\documentclass[12pt]{article}
\usepackage{amsfonts}
\usepackage{amsmath, amsthm}
\usepackage{epsfig}
\usepackage{algorithm}

\usepackage{soul}
\usepackage[normalem]{ulem}
\usepackage{amsfonts}
\usepackage{subcaption}
\usepackage{epsfig}
\usepackage{booktabs}
\usepackage{lscape}
\usepackage{multirow}

\usepackage{longtable}
\usepackage{rotating}
\usepackage{array}
\usepackage{multicol}
\usepackage{graphicx}
\usepackage{bigstrut}
\usepackage{setspace}
\usepackage{epstopdf}
\usepackage{mathrsfs}
\usepackage{amsfonts}
\usepackage{epsfig}
\usepackage{booktabs}
\usepackage{threeparttable}
\usepackage{epstopdf}
\usepackage{lscape}
\usepackage{multirow}

\usepackage{algorithmic}
\usepackage{amsmath,verbatim,color,amssymb,epsfig}
\usepackage{bm}
\usepackage{amsfonts}
\usepackage{epsfig}
\usepackage{multirow}
\usepackage{graphicx}
\usepackage{array}
\usepackage[table]{xcolor}
\definecolor{Gray}{gray}{0.80}
\usepackage{lscape}
\usepackage{cite}
\usepackage{natbib}
\usepackage[normalem]{ulem}
\usepackage[colorlinks=true,urlcolor=blue,citecolor=black,linkcolor=blue,bookmarks=true]{hyperref}
\usepackage{caption}
\begin{document}
\def\eqx"#1"{{\label{#1}}}
\def\eqn"#1"{{\ref{#1}}}

\makeatletter 
\@addtoreset{equation}{section}
\makeatother  

\def\yincomment#1{\vskip 2mm\boxit{\vskip 2mm{\color{red}\bf#1} {\color{blue}\bf --Yin\vskip 2mm}}\vskip 2mm}
\def\squarebox#1{\hbox to #1{\hfill\vbox to #1{\vfill}}}
\def\boxit#1{\vbox{\hrule\hbox{\vrule\kern6pt
          \vbox{\kern6pt#1\kern6pt}\kern6pt\vrule}\hrule}}

\newcommand{\blue}[1]{\textcolor{blue}{{#1}}}
\newcommand{\red}[1]{\textcolor{red}{{#1}}}
\def\theequation{\thesection.\arabic{equation}}
\newcommand{\ds}{\displaystyle}

\newcommand{\bJ}{\mbox{\bf J}}
\newcommand{\bF}{\mbox{\bf F}}
\newcommand{\bM}{\mbox{\bf M}}
\newcommand{\bR}{\mbox{\bf R}}
\newcommand{\bZ}{\mbox{\bf Z}}
\newcommand{\bX}{\mbox{\bf X}}
\newcommand{\bx}{\mbox{\bf x}}
\newcommand{\bQ}{\mbox{\bf Q}}
\newcommand{\bH}{\mbox{\bf H}}
\newcommand{\bh}{\mbox{\bf h}}
\newcommand{\bz}{\mbox{\bf z}}
\newcommand{\ba}{\mbox{\bf a}}
\newcommand{\be}{\mbox{\bf e}}
\newcommand{\bG}{\mbox{\bf G}}
\newcommand{\bB}{\mbox{\bf B}}
\newcommand{\bb}{\mbox{\bf b}}
\newcommand{\bA}{\mbox{\bf A}}
\newcommand{\bC}{\mbox{\bf C}}
\newcommand{\bI}{\mbox{\bf I}}
\newcommand{\bD}{\mbox{\bf D}}
\newcommand{\bU}{\mbox{\bf U}}
\newcommand{\bc}{\mbox{\bf c}}
\newcommand{\bd}{\mbox{\bf d}}
\newcommand{\bs}{\mbox{\bf s}}
\newcommand{\bS}{\mbox{\bf S}}
\newcommand{\bV}{\mbox{\bf V}}
\newcommand{\bv}{\mbox{\bf v}}
\newcommand{\bW}{\mbox{\bf W}}
\newcommand{\bw}{\mbox{\bf w}}
\newcommand{\bg}{\mbox{\bf g}}
\newcommand{\bu}{\mbox{\bf u}}
\def\bb{{\bf b}}

\newcommand{\bcU}{\boldsymbol{\cal U}}
\newcommand{\bbeta}{\boldsymbol{\beta}}
\newcommand{\bdelta}{\boldsymbol{\delta}}
\newcommand{\bDelta}{\boldsymbol{\Delta}}
\newcommand{\boldeta}{\boldsymbol{\eta}}
\newcommand{\bxi}{\boldsymbol{\xi}}
\newcommand{\bGamma}{\boldsymbol{\Gamma}}
\newcommand{\bSigma}{\boldsymbol{\Sigma}}
\newcommand{\balpha}{\boldsymbol{\alpha}}
\newcommand{\bOmega}{\boldsymbol{\Omega}}
\newcommand{\btheta}{\boldsymbol{\theta}}
\newcommand{\bmu}{\boldsymbol{\mu}}
\newcommand{\bnu}{\boldsymbol{\nu}}
\newcommand{\bgamma}{\boldsymbol{\gamma}}

\newtheorem{thm}{Theorem}[section]
\newtheorem{lem}{Lemma}[section]
\newtheorem{rem}{Remark}[section]
\newtheorem{cor}{Corollary}[section]
\newcolumntype{L}[1]{>{\raggedright\let\newline\\\arraybackslash\hspace{0pt}}m{#1}}
\newcolumntype{C}[1]{>{\centering\let\newline\\\arraybackslash\hspace{0pt}}m{#1}}
\newcolumntype{R}[1]{>{\raggedleft\let\newline\\\arraybackslash\hspace{0pt}}m{#1}}

\newcommand{\tabincell}[2]{\begin{tabular}{@{}#1@{}}#2\end{tabular}}

\baselineskip=24pt
\begin{center}
{\Large \bf
Demystify Lindley's Paradox by Interpreting $P$-value
as Posterior Probability
}
\end{center}

\vspace{2mm}
\begin{center}
{\bf Guosheng Yin$^{1}$ and Haolun Shi$^{2}$}
\end{center}

\begin{center}

$^{1}$Department of Statistics and Actuarial Science\\
The University of Hong Kong\\
Pokfulam Road, Hong Kong\\

\vspace{2mm}

$^{2}$Department of Statistics and Actuarial Science\\
School of Computing Science\\
Simon Fraser University \\
Burnaby, BC, Canada\\

$^{1}${\em Correspondence email}: gyin@hku.hk\\


\end{center}

\noindent{Abstract.} In the hypothesis testing framework,
$p$-value is often computed 
to determine rejection of the null hypothesis or not.
On the other hand, Bayesian approaches typically compute the posterior probability
of the null hypothesis to evaluate its plausibility.
We revisit Lindley's paradox (Lindley, 1957) and demystify the conflicting
results between Bayesian and frequentist hypothesis testing
procedures by casting a two-sided hypothesis as
a combination of two one-sided hypotheses along the opposite directions.
This can naturally circumvent the ambiguities of assigning a point mass
to the null and choices of using local or non-local prior
distributions. As $p$-value solely depends on
the observed data without incorporating any prior information, 
we consider non-informative prior distributions for fair comparisons with $p$-value.
The equivalence of $p$-value and the Bayesian posterior
probability of the null hypothesis can be established to
reconcile Lindley's paradox.
Extensive simulation studies are conducted with 
multivariate normal data and random effects models 
to examine the relationship between the $p$-value and
posterior probability. 

\vspace{0.5cm}
\noindent{KEY WORDS:} Bayesian posterior
probability, Hypothesis testing, Interpretation of $p$-value,
Point null hypothesis, Two-sided test

\vspace{1.5cm}
\section{Introduction}
Frequentist hypothesis testing is widely used in scientific
studies and computation of $p$-value is one of the critical
components in the testing procedure.
The $p$-value is defined as the probability of observing the
random data as or more extreme than the observed
given the null hypothesis being true.
By setting the statistical significance level at 5\%,  
a $p$-value smaller than 5\% is considered statistically significant
which leads to rejection of the null hypothesis,
and that greater than 5\% is
considered statistically insignificant which results
in failure to reject the null. However, $p$-value has been
largely criticized for its misuse or misinterpretations, and oftentimes
it is recommended to resort to Bayesian methods, 
such as the posterior probability of
the null/alternative hypothesis and Bayes factor.  For example,
Goodman (1999) supports the
Bayes factor in contrast to the $p$-value as a measure of evidence
in medical research. In psychology research,
Wagenmakers (2007) reveals the issues with $p$-values and
recommends use of the Bayesian information criterion instead,
and Hubbard and Lindsay (2008) claim that $p$-values
tend to exaggerate the evidence against the null hypothesis.

Extensive research work have been conducted to
reconcile the differences between Bayesian and frequentist analysis
(Pratt, 1965; Berger, 2003; and Bayarri and Berger, 2004). Frequentist
methods do not rely upon any prior information but the observed
data, and thus for a fair comparison, non-informative prior should be used
in Bayesian analysis although one major advantage of Bayesian
approaches is to incorporate prior information in a natural way.
Particularly, Berger and Sellke (1987), Berger and Delampady (1987), and
Casella and Berger (1987) investigate the relationships between the $p$-value
and Bayesian measure of evidence against the null hypothesis for hypothesis testing.
Sellke, Bayarri, and Berger (2001) propose to calibrate
$p$-values for testing precise null hypotheses.


More recently, extensive discussions on modern statistical inference in a special issue of {\em The American Statistician} highlight several insights regarding the role of $p$-value and Bayesian statistics. In addition to the usual criticisms on null hypothesis significance testing
(McShane et al., 2019; Wasserstein et al., 2019)
and recommendations for improving the use of $p$-value
for statistical inference (Benjamin and Berger, 2019; Matthews, 2019; Betensky, 2019),
of particular interest are a collection of articles on the connection of the statistical significance under frequentist inference to the Bayesian paradigm, as well as various Bayesian alternatives to $p$-value.
Ioannidis (2019) investigates the abuse of $p$-value in the scientific literature and presents several alternatives to $p$-value such as confidence intervals, false discovery rates, and Bayesian methods.
Gannon et al. (2019) propose a testing procedure
based on a mixture of frequentist and Bayesian tools.
Kennedy-Shaffer (2019) contrasts the frequentist and Bayesian inferential
frameworks from a historical perspective.
Rougier (2019) shows that under certain context, the $p$-value is never greater than the Bayes factor through an inequality based on the generalized likelihood ratio.
Johnson (2019) compares the likelihood ratio test and Bayes factor in the context of a marginally significant $t$-test and suggests a more stringent standard of evidence.
Billheimer (2019) proposes a new method for statistical inference based on Bayesian predictive distributions.
Colquhoun (2019) reflects on the status quo of the misuse of $p$-value and suggests converting the observed $p$-value to the Bayesian false positive risk.
Krueger and Heck (2019) recommend using $p$-value as a heuristic guide for estimating the posterior probability of the null.
Manski (2019) proposes to use the Bayesian decision theory as an aid for treatment selection in medical studies, and Ruberg et al. (2019) present several practical applications of Bayesian methods.

There are often ambiguities on prior specification with the point null
and composite alternative hypotheses in the Bayesian paradigm
(Casella and Berger, 1987; and Johnson and Rossell, 2010).
Under non-informative priors, Shi and Yin (2020) interpret
$p$-value as the posterior probability of the null hypothesis
under both one- and two-sided hypothesis tests.
We revisit Lindley's paradox and 
for the point null hypothesis in a two-sided test we reformulate
the problem as a combination of two one-sided null hypotheses.
As a result, the ambiguities on prior specification disappear, and
this gives a new explanation to reconcile the
differences between Bayesian and frequentist approaches.

The rest of the paper is organized as follows. In Section 2,
we present a motivating example to demonstrate
how a point null hypothesis in a two-sided test can be
reformulated as a combination of two one-sided tests, which
naturally reconcile Lindley's paradox. In Section 3,
we revisit Lindley's original paradox and show that the $p$-value and the posterior
probability of the null have an equivalence relationship
under non-informative priors.
Section 4 considers hypothesis testing with normal data,
and Section 5 extends the result to multivariate
tests. We develop similar results for hypothesis
testing of variance components under random effects models in Section 6.
Finally, Section 7 concludes with some remarks.

\section{Motivating Example}
\subsection{Illustration of Lindley's Paradox}
In the hypothesis testing framework, it may happen that
the Bayesian and frequentist approaches produce
opposite conclusions for certain choices of the prior distribution
(e.g., the witch hat prior---a point mass at the null and flat elsewehere).
To illustrate Lindley's paradox (Lindley, 1957),
we start with a simple example. Suppose that 28,298 boys and 27,801 girls
were born in a city last year. The observed proportion of male
births in the city is $y=28298/56099\approx 0.5044297$.
Let $\theta$ denote the true proportion
of male births, and we are interested in testing
\begin{equation*}\label{twosidedH}
H_0: \theta=0.5 \quad {\rm versus} \quad H_1: \theta \not=0.5.
\end{equation*}

\subsubsection{$p$-value from an exact test}
The number of male births follows a binomial distribution with mean
$n\theta$ and variance $n\theta(1-\theta)$, where $n=56,099$ is the
total number of births.
Under the frequentist paradigm, the $p$-value based on the binomial exact test is
\begin{equation*}
\Pr(Y\ge y| H_0)= \sum_{x=28298}^{n}\binom{n}{x}0.5^{n}
\approx  0.01812363.
\end{equation*}

\subsubsection{$p$-value using normal approximation}
On the other hand, as the sample size $n$ is large and the
observed male proportion $y$ is not close to 0 or 1, we can use
normal approximation to simplify the computation, so we assume
$Y\sim N(\theta, \hat\sigma^2)$ where $\hat\sigma^2 = y(1-y)/n$.
The frequentist approach calculates the $p$-value
as the upper tail probability of as or more extreme than
the observed data under the null distribution,
\begin{equation}\label{pvaluebirth}
\Pr(Y\ge y| H_0)= \int_{28298/56099}^\infty
\frac{1}{\sqrt{2\pi}\hat\sigma}\exp\bigg\{-\frac{(x-0.5)^2}{2\hat\sigma^2} \bigg\}d x
\approx  0.01793329.
\end{equation}
Evidently, the exact and approximate $p$-values are very close.
As the hypothesis test is two-sided, the final $p$-value is $2\times 0.01793329
\approx  0.03586658$.
At the typical significance level of 5\%, we clearly reject $H_0$.

\subsubsection{Bayesian posterior probability of $H_0$}
If we proceed with a Bayesian approach, the usual approach is to first
specify a prior distribution on $H_0$ and $H_1$.  Without any preference,
we assign an equal prior probability to $H_0$ and $H_1$,
i.e., $P(H_0)=P(H_1)=0.5$. Under $H_0$,
$\theta$ has a point mass at 0.5. Under $H_1$, $\theta$ is not equal to
0.5 and, to be fair, we assign a uniform prior distribution to $\theta$
on $[0, 1]$. As a result, the posterior probability of $H_0$ is
\begin{eqnarray*}
P(H_0 | y)&=&\frac{P(y|H_0)P(H_0)}{P(y|H_0)P(H_0)+P(y|H_1)P(H_1)}\\
&=&\frac{\ds \exp\bigg\{-\frac{(y-0.5)^2}{2\hat\sigma^2}\bigg\}}
{\ds \exp\bigg\{-\frac{(y-0.5)^2}{2\hat\sigma^2}\bigg\}
+\int_{0}^1 \exp\bigg\{-\frac{(y-\theta)^2}{2\hat\sigma^2}\bigg\}
d \theta }\\
&\approx&  0.9543474,
\end{eqnarray*}
which strongly supports $H_0$.

Such conflict between
Bayesian and frequentist hypothesis testing approaches
may happen when the prior distribution is a mixture of
a sharp peak at $H_0$ and no sharp features anywhere else,
which is often known as Lindley's paradox.
We explain as follows that such a conflicting
result can be resolved if we view the two-sided hypothesis as a combination
of two one-sided hypotheses, and further demonstrate the equivalence
of $p$-value and the posterior probability of the null when a non-informative
prior is used.

\subsection{One-sided Hypothesis Test}
For ease of exposition, we start with a one-sided hypothesis test,
$$H_0: \theta \le 0.5 \quad {\rm versus} \quad H_1: \theta > 0.5.$$
The $p$-value is still calculated in the same way,
as the upper tail probability of as or more extreme than
the observed data under the null distribution.
Under the normal approximation, following (\ref{pvaluebirth}),
$p$-value $=0.01793329$.

\subsubsection{Using Bayes' theorem}
In the Bayesian approach,
we assign a uniform prior distribution to $\theta$, i.e.,
$\theta\sim {\rm Unif}[0, 1]$, so the prior probabilities $P(H_0)=P(H_1)=1/2$.
Under normal approximation, the posterior probability of $H_0$ is
\begin{eqnarray}\label{e1}
P(H_0 | y)&=&\frac{P(y|H_0)P(H_0)}{P(y|H_0)P(H_0)+P(y|H_1)P(H_1)}\nonumber\\
&=&\frac{\ds \int_{0}^{0.5} \exp\bigg\{-\frac{(y-\theta)^2}{2\hat\sigma^2}\bigg\}d \theta}
{\ds \int_{0}^{0.5}\exp\bigg\{-\frac{(y-\theta)^2}{2\hat\sigma^2}\bigg\}d \theta
+\int_{0.5}^1 \exp\bigg\{-\frac{(y-\theta)^2}{2\hat\sigma^2}\bigg\}
d \theta }\nonumber\\
&\approx& 0.01793329.
\end{eqnarray}
which is the same as the $p$-value in (\ref{pvaluebirth}).

\subsubsection{Using the posterior distribution of the parameter}
Under the normal approximation,
an alternative way is to first obtain the posterior distribution of $\theta$,
by assuming the prior distribution of $\theta$ to be flat, i.e., $p(\theta)\propto 1$.
The posterior distribution of $\theta$ is then given by
$$ P(\theta | y) \propto \exp\bigg\{-\frac{(\theta-\hat{\theta})^2}{2\hat\sigma^2}\bigg\},$$
i.e., $\theta | y \sim N(\hat{\theta}, \hat\sigma^2)$ where $\hat{\theta}=y$.
As a result, we can compute
\begin{equation}\label{e2}
P(H_0 | y)= P(\theta \le 0.5 | y)= \int_{-\infty}^{0.5}
\frac{1}{\sqrt{2\pi}\hat\sigma}\exp
\bigg\{-\frac{(\theta-28298/56099)^2}{2\hat\sigma^2} \bigg\}d\theta = p\mbox{-value},
\end{equation}
which is exactly the same as the $p$-value in (\ref{pvaluebirth}), because it is
easy to show that
$$\int_{-\infty}^{a}
\frac{1}{\sqrt{2\pi}\sigma}\exp
\bigg\{-\frac{(x-b)^2}{2\sigma^2} \bigg\}dx
=\int_{b}^{\infty}
\frac{1}{\sqrt{2\pi}\sigma}\exp
\bigg\{-\frac{(x-a)^2}{2\sigma^2} \bigg\}dx,
$$
for any values of $a$ and $b$ on the real line.

\subsubsection{Bayesian exact beta distribution}
If we do not assume the asymptotic normal distribution, we
can proceed with Bayesian exact computation.
Under the Bayesian paradigm, if we assume a uniform prior for $\theta$,
i.e., $\theta \sim {\rm Beta}(1,1)$,
the posterior distribution of $\theta$ is still Beta, i.e.,
$\theta | y \sim {\rm Beta}(ny+1,n-ny+1)$.
The posterior probability of the null can be directly calculated as
\begin{equation*}
\Pr(H_0|y) =
\int_0^{0.5} \frac{\Gamma(n+2)}{\Gamma(ny+1)\Gamma(n-ny+1)}\theta^{ny}(1-\theta)^{n-ny} d\theta
\approx 0.01793728,
\end{equation*}
which is close to the $p$-value. Note that this procedure
does not use the normal approximation. We further experiment other
non-informative Beta prior distribution by choosing
$\theta \sim {\rm Beta}(\alpha,\beta)$ with
$\alpha=\beta=0.1, 0.01, 0.001, 0.0001, 0.00001, 0.000001$, and
the result is given in Table \ref{tbl1}. Clearly, under
non-informative prior distributions, the posterior probabilities
of the null are very close to the $p$-value.

\subsection{Two-sided Hypothesis Test}
In a two-sided hypothesis test, the prior specification on the point
null is often ambiguous by assigning a point probability mass.
To circumvent the issue of point mass,
we rewrite the two-sided hypothesis in (\ref{twosidedH}) as a combination of
two one-sided hypotheses:
\begin{equation}\label{twoH}
\left\{\begin{array}{ll}
H_0: & \theta\le 0.5 \quad {\rm versus} \quad H_1: \theta > 0.5,\\
H_0: & \theta\ge 0.5 \quad {\rm versus} \quad H_1: \theta < 0.5.
\end{array}
\right.
\end{equation}

Under the frequentist paradigm, the $p$-value for the first one-sided hypothesis test
in (\ref{twoH}),
$H_0: \theta\le 0.5 \ {\rm versus} \ H_1: \theta > 0.5,$
is given by
\begin{equation*}
 \Pr(Y\ge y| H_0)= 1 - \Phi({28298/56099};0.5,\hat\sigma^2)
\approx  0.01793329,
\end{equation*}
where $\Phi(\cdot;\mu,\hat\sigma^2)$ denotes the
cumulative distribution function (CDF) of a normal random variable
with mean $\mu$ and variance $\hat\sigma^2$.
The $p$-value for the second one-sided hypothesis test in (\ref{twoH}),
$H_0: \theta\ge 0.5 \ {\rm versus} \ H_1: \theta < 0.5,$
is given by
\begin{equation*}
 \Pr(Y\le y| H_0)= \Phi({28298/56099};0.5,\hat\sigma^2)
\approx  0.9820667.
\end{equation*}
Therefore, the $p$-value under the two-sided hypothesis test in (\ref{twoH}) is
given by
\begin{equation*}
 p \mbox{-value}_2 =  2\times {\rm min}\{\Pr(Y\le y| H_0),\Pr(Y\ge y| H_0) \}
 =  2\times 0.01793329=0.03586658.
\end{equation*}

As a counterpart,
we propose a new concept of the two-sided posterior probability (${\rm PoP}_2$),
defined as
\begin{eqnarray*}
{\rm PoP_2} &=& 2\times {\rm min}\{\Pr({\theta} \le 0.5|y), \Pr({\theta} \ge 0.5|y)\}\\
&=& 2\times {\rm min}\{0.01793329, 0.9820667\}\\
&=& 0.03586658.
\end{eqnarray*}
Therefore, it is evident that the value of ${\rm PoP_2}$ is the same as the
two-sided hypothesis testing $p$-value under normal approximation.
If an equal prior
probability is assumed for $H_0$ and $H_1$, then the Bayes factor in favor of $H_0$
over $H_1$, denoted as ${\rm BF}_{0,1}$ can be calculated as the odds of
the $p$-value,
$$ {\rm BF}_{0,1}=\frac{p\mbox{-value}}{1-p\mbox{-value}}.$$

\section{Lindley's Paradox}
It is well-known that Bayesian methods adhere to the likelihood principle; that is,
all that we know about the data or the sample
is contained in the likelihood function. If the likelihood functions under
two different sampling plans or sampling distributions
are proportional with respect to the parameter of interest $\theta$,
statistical inferences on $\theta$ should be identical
based on these two sampling distributions. However, frequentist
approaches may result in two different conclusions in
the hypothesis testing framework.

\subsection{Original Coin-tossing Example}
We consider an experiment in which a coin was tossed
12 times, with 9 heads and 3 tails observed
(Lindley and Phillips, 1976). Let $\theta$ be the
probability of observing a head for a toss of the coin,
and we are interested in testing the hypotheses,
$$H_0\mbox{:} \  \theta = 0.5\quad {\rm versus}\quad H_1\mbox{:} \  \theta > 0.5.$$
There is no further information on the sampling plan.

Based on the observed data, there could be two choices for the likelihood
function. First, let $Y$ denote the number of heads after a fixed number
of $n$ tosses; that is, $Y\sim \mbox{Bin}(n, \theta)$.
Under the binomial distribution
with $n=12$ tosses and $y=9$ heads observed,
the likelihood function is given by
$$ L_{\rm B}(\theta | y)=\binom{n}{y}\theta^y(1-\theta)^{n-y}=
\binom{12}{9}\theta^9(1-\theta)^3.$$
Second, let $Y$ be the number of heads for the
tosses of the coin until the third tail $(r=3)$ is observed; that is,
$Y\sim \mbox{Neg-Bin}(r, \theta)$.
Under the negative binomial distribution,
the likelihood function is given by
$$ L_{\rm NB}(\theta | y)=\binom{y+r-1}{y}\theta^y(1-\theta)^{r}=
\binom{11}{9}\theta^9(1-\theta)^3.$$
Clearly, the two likelihood functions are proportional
to each other up to a
normalizing constant, i.e.,
$L_{\rm B}(\theta | y)\propto L_{\rm NB}(\theta | y)$. As a result, the posterior
distributions of $\theta$ under these two sampling distributions
are identical in the Bayesian framework. However, frequentist
inferences about $\theta$ are very different, which depends on
the sampling distribution. In particular, we can calculate
the $p$-value, which is the probability of obtaining
the result as or more extreme than the observed
assuming that $H_0$ is true.
Based on the binomial likelihood, the $p$-value is
$$ p\mbox{-value}_{\rm B}=\Pr(y\ge 9| H_0)=
\sum_{y=9}^{12}\binom{12}{y}0.5^{12}\approx 0.07299805,$$
while under the negative binomial distribution,
$$ p\mbox{-value}_{\rm NB}=\Pr(y\ge 9| H_0)=
\sum_{y=9}^\infty\binom{y+2}{y}0.5^{3+y}\approx 0.03271484.$$
If we set the significance level at $\alpha=0.05$, the frequentist
hypothesis test yields conflicting results: The null hypothesis
is accepted under the binomial distribution, but it is rejected
under the negative binomial distribution.

\subsection{One-sided Hypothesis Test}
Suppose that we conduct a one-sided hypothesis test,
\[{H_0}:{\theta} \le 0.5 \quad {\rm versus}\quad {H_1}:{\theta} > 0.5. \]
Under the Bayesian paradigm, if
we assume a symmetric beta prior distribution $(\alpha=\beta)$ for $\theta$,
i.e., $\theta\sim {\rm Beta}(\alpha,\beta)$, then
the posterior distribution of $\theta$ is Beta$(y+\alpha,n-y+\beta)$.
The posterior probability of the null can be computed as
\begin{equation}\label{bayes}
\Pr(H_0|y) =
\int_0^{0.5} \frac{\Gamma(n+\alpha+\beta)}{\Gamma(n-y+\beta)\Gamma(y+\alpha)}
\theta^{y+\alpha-1}(1-\theta)^{n-y+\beta-1} d\theta
\end{equation}

\begin{figure}[htbp]
\setlength{\abovecaptionskip}{-5pt}
\setlength{\belowcaptionskip}{5pt}
\begin{center}
\includegraphics[angle=0,width=2.7 in, totalheight=2.7 in]{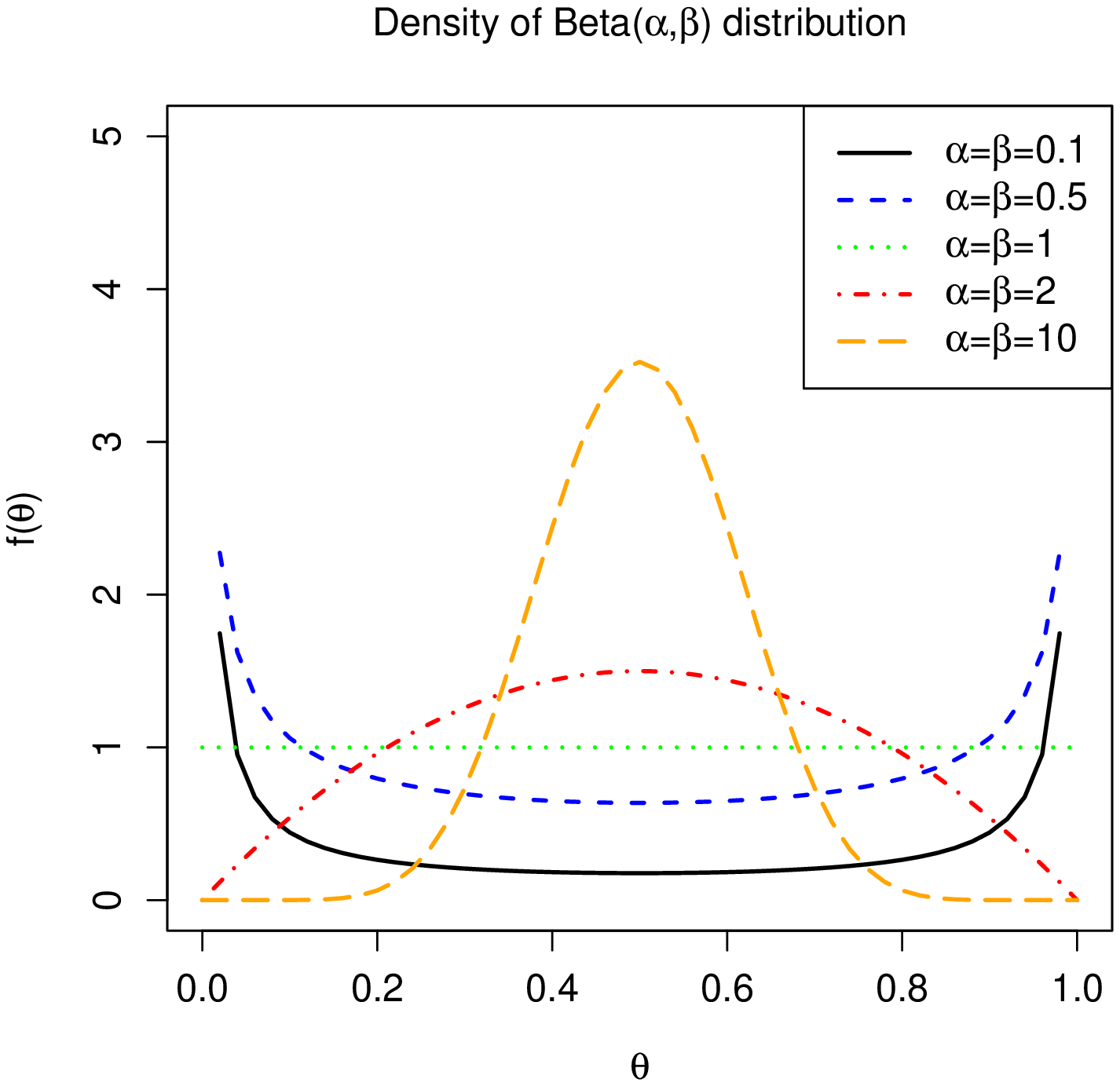}\\
\includegraphics[angle=0,width=2.7 in, totalheight=2.7 in]{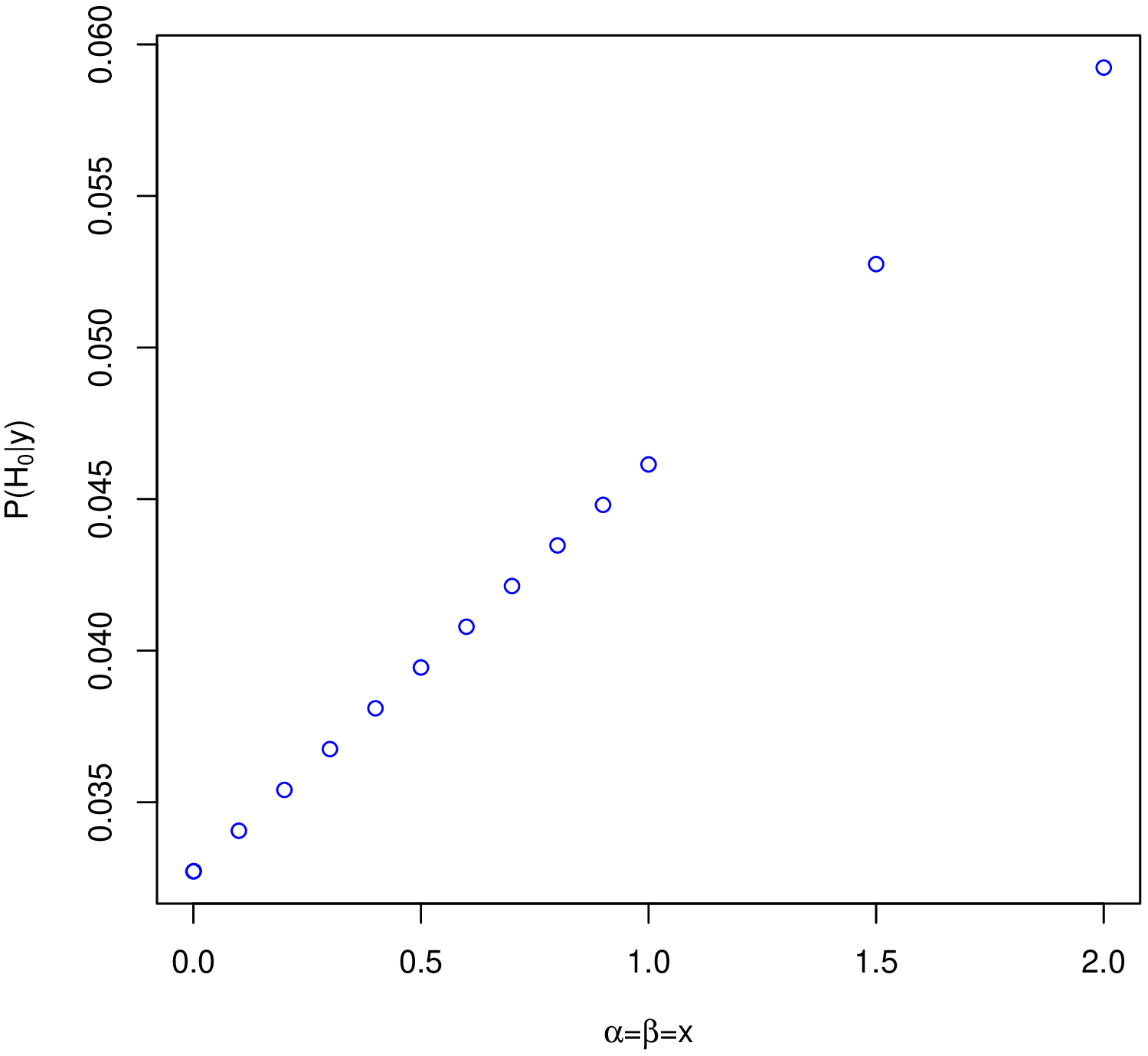}\\
\includegraphics[angle=0,width=2.7 in, totalheight=2.7 in]{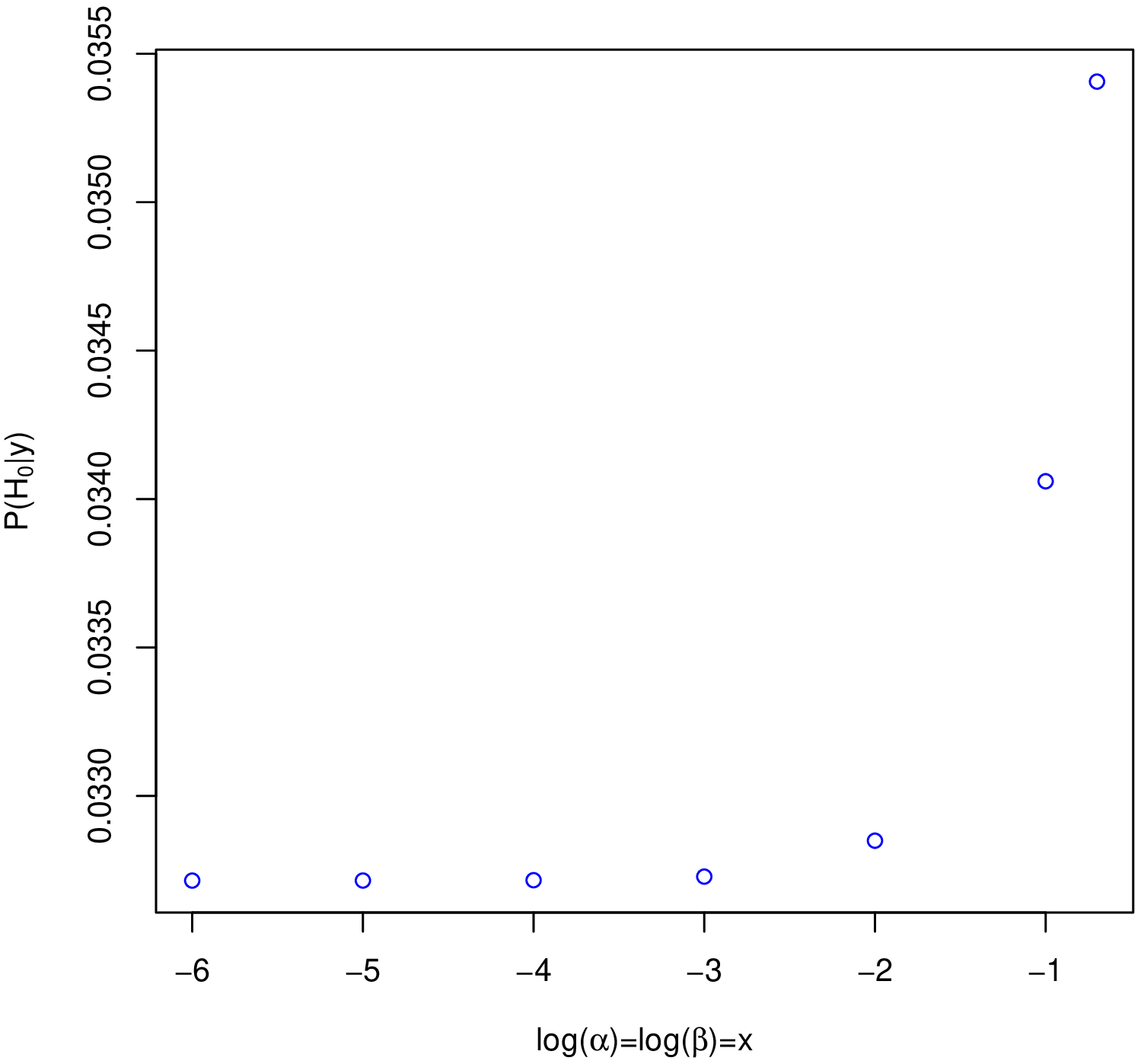}
\end{center}
\caption{\label{Fig:postprobh0}
The Beta$(\alpha, \beta)$ prior
distribution with $\alpha=\beta$ (top panel);
the posterior probability of $H_0$
under different Beta$(\alpha, \beta)$ prior distributions with $\alpha=\beta$ (middle panel);
the zoom-in plot in the corner $(0, 0)$ of the middle panel
by taking the log transformation of the Beta prior hyperparameters (bottom panel).}
\end{figure}

The top panel of
Figure \ref{Fig:postprobh0} shows the different beta prior distributions
Beta$(\alpha, \beta)$ with $\alpha=\beta$, and the middle panel
exhibits the pattern of
the posterior probability of $H_0$ under different hyperparameter values $\alpha=\beta$
from $10^{-6}$ to 2.
Under such prior distributions, the implicit probability of landing
on a head for a coin toss is 0.5, which is
smaller than the one observed in the actual data, $9/12=0.75$.
When the value of $\alpha=\beta$ increases, the prior distribution
becomes more centered at the null value 0.5. As the information in the prior distribution
strengthens, the prior plays an increasingly important role in the posterior distribution,
so that the posterior probability of $H_0$ increases under the influence of the
strengthening prior information. The bottom panel in Figure \ref{Fig:postprobh0} shows
the zoom-in plot in the corner $(0,0)$
of the top panel by taking the log transformation of
the x-axis. Table \ref{popnull} shows the values of the posterior probability $P(H_0|y)$
for different values of the hyperparameters in the Beta$(\alpha,\beta)$ prior
distribution with $\alpha=\beta$. The conclusion is that as the values of
$\alpha=\beta$ decrease toward zero, i.e., the prior becomes less and less
informative, $P(H_0|y)$ approaches the $p$-value obtained
from the negative binomial distribution.

\subsection{Equivalence Between the Negative Binomial $P$-value and
the Posterior Probability of the Null}
The CDF of a negative binomial distribution, $\mbox{Neg-Bin}(r, \theta)$, is
denoted as
$$F_{\rm NB}(y;r,\theta) = 1 - I_{\theta}(y+1,r),$$
where $I_x(a,b)$ is the regularized incomplete beta function defined as
$$
I_x(a,b) = \frac{B(x;a,b)}{B(a,b)},
$$
with
\begin{align*}
B(x;a,b) =& \int_{0}^{x} t^{a-1}(1-t)^{b-1}dt, \\
B(a,b) =& \int_{0}^{1}t^{a-1}(1-t)^{b-1} dt.
\end{align*}
Therefore, the $p$-value based on the assumption
$Y\sim \mbox{Neg-Bin}(r= n - y, \theta)$ is
\begin{equation}\label{pvaluenb}
p\mbox{-value}_{\rm NB}=\Pr(Y \ge y| H_0)= 1 - F_{\rm NB}(y-1;r,\theta=0.5) = I_{0.5}(y,r)=I_{0.5}(y,n-y).
\end{equation}

Under the Bayesian paradigm, if we assume a Beta$(\alpha,\beta)$ prior
distribution for $\theta$,
the posterior distribution of $\theta$ is Beta$(y+\alpha,n-y+\beta)$.
The CDF of a Beta$(a,b)$ distribution is
$F_{\rm Beta}(x;a,b)  = I_{x}(a,b).$
Hence, the posterior probability of the null is
\begin{equation}\label{pop}
P(H_0 | y)=  F_{\rm Beta}(0.5;y+\alpha,n-y+\beta) = I_{0.5}(y+\alpha,n-y+\beta).
\end{equation}
Comparing (\ref{pvaluenb}) and (\ref{pop}),
when the hyperparameters $\alpha$ and $\beta$ are very small relative to $n$ and $y$,
the $p$-value under the negative binomial model is close to the posterior probability of the null.

\subsection{Numerical Study}
\begin{figure}[htbp]
\setlength{\abovecaptionskip}{-5pt}
\setlength{\belowcaptionskip}{5pt}
\begin{center}
\includegraphics[angle=0,width=2.7 in, totalheight=2.7 in]{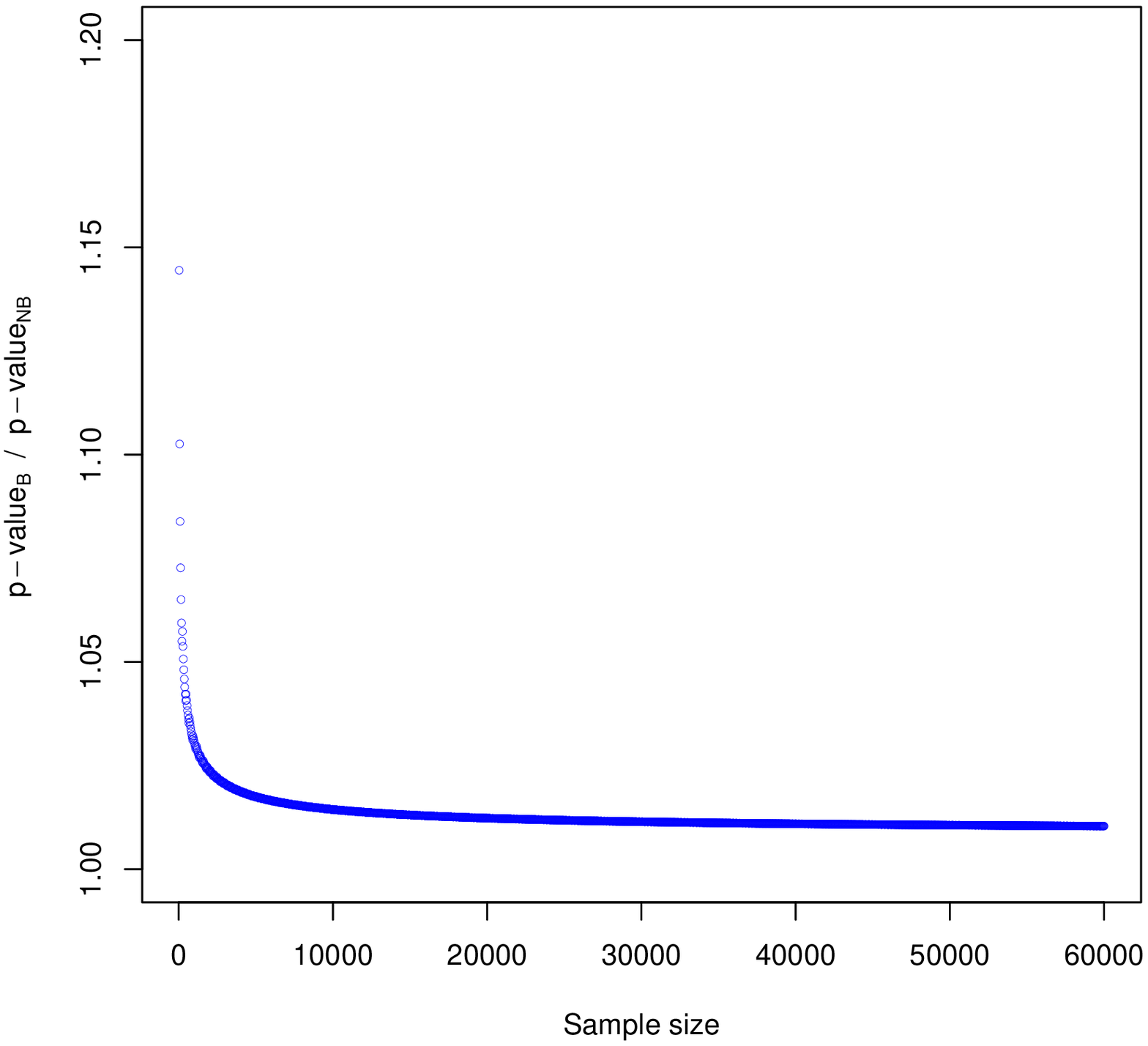}
\includegraphics[angle=0,width=2.7 in, totalheight=2.7 in]{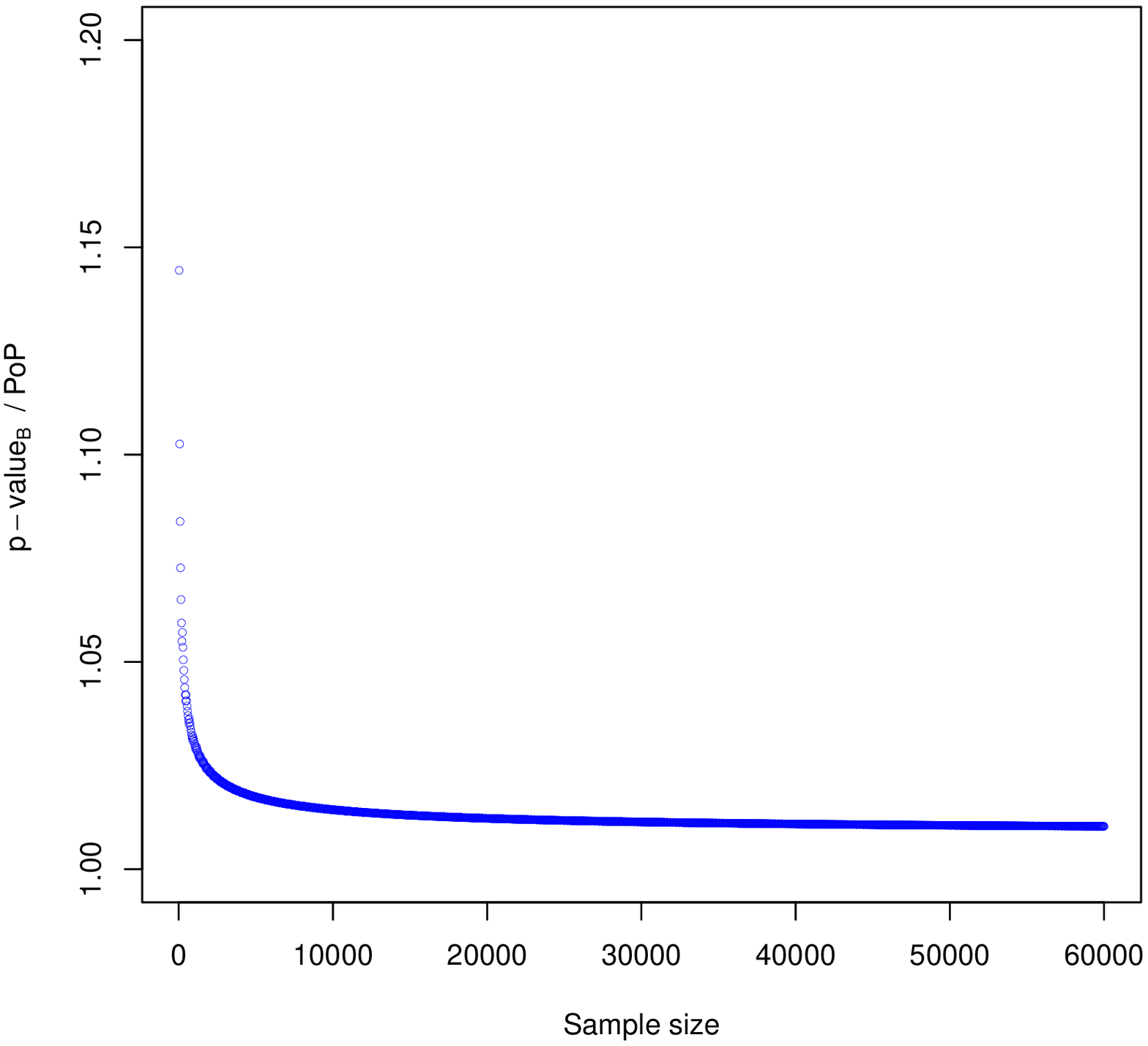}\\
\includegraphics[angle=0,width=2.7 in, totalheight=2.7 in]{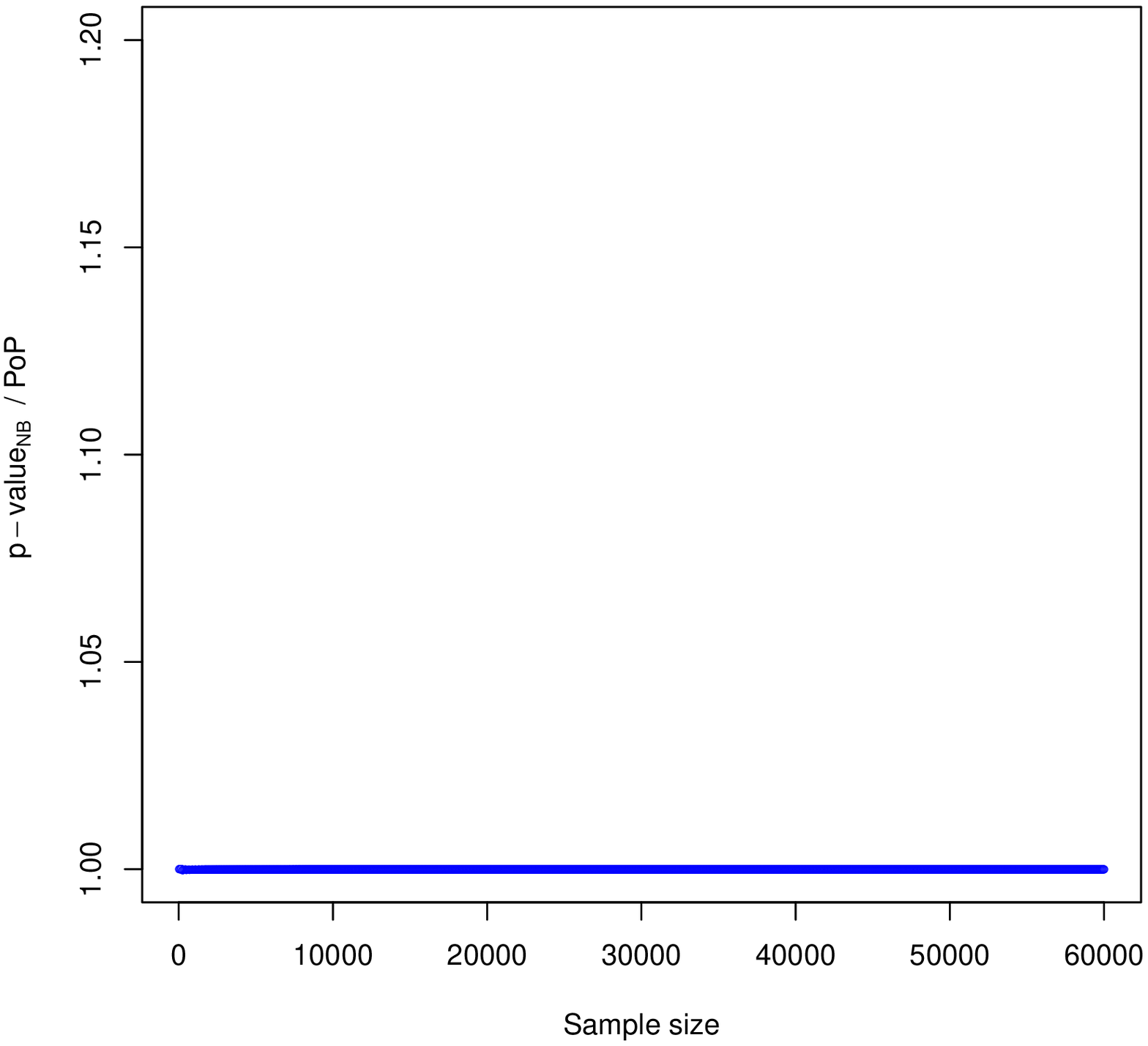}
\includegraphics[angle=0,width=2.7 in, totalheight=2.7 in]{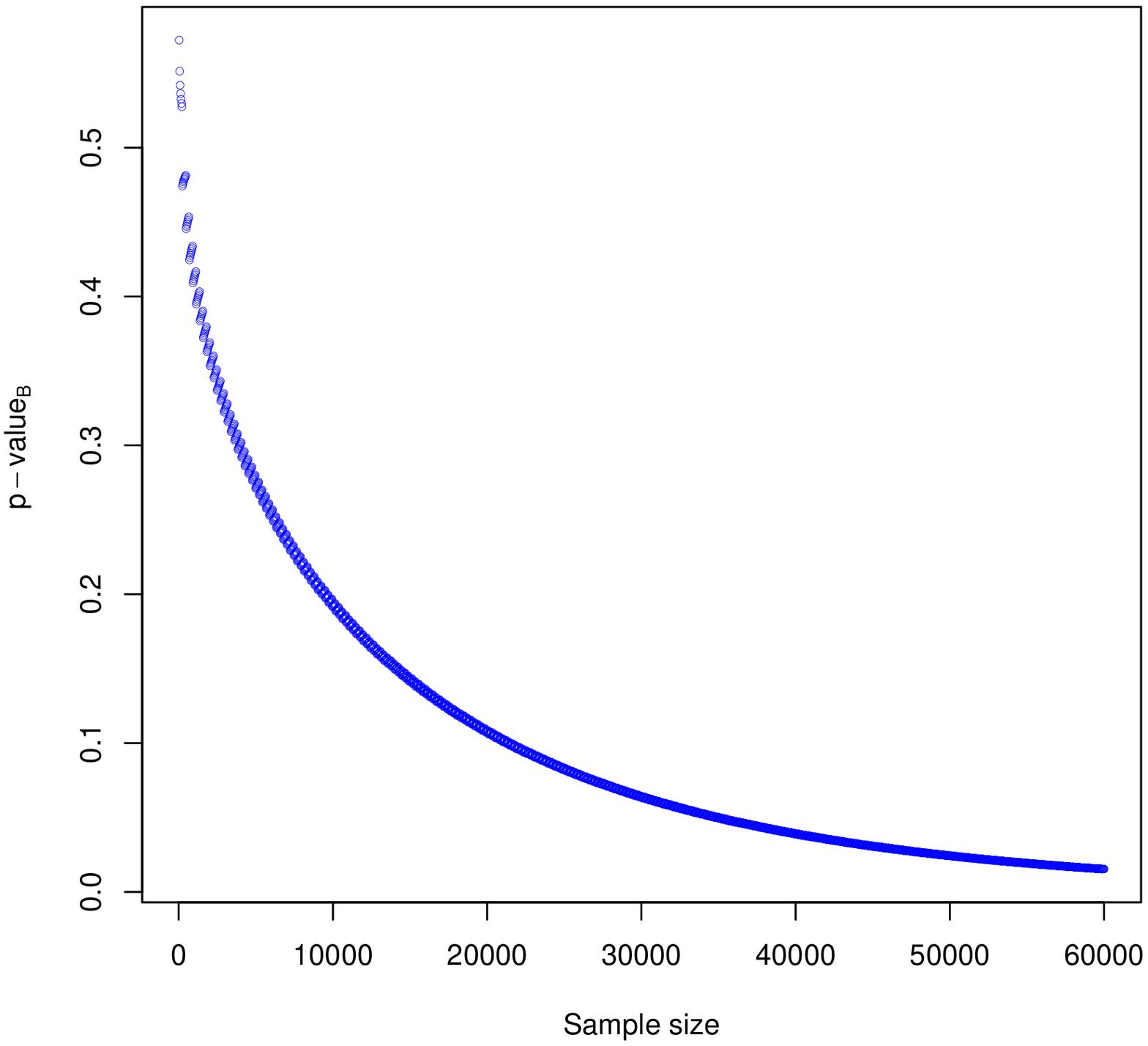}\\
\includegraphics[angle=0,width=2.7 in, totalheight=2.7 in]{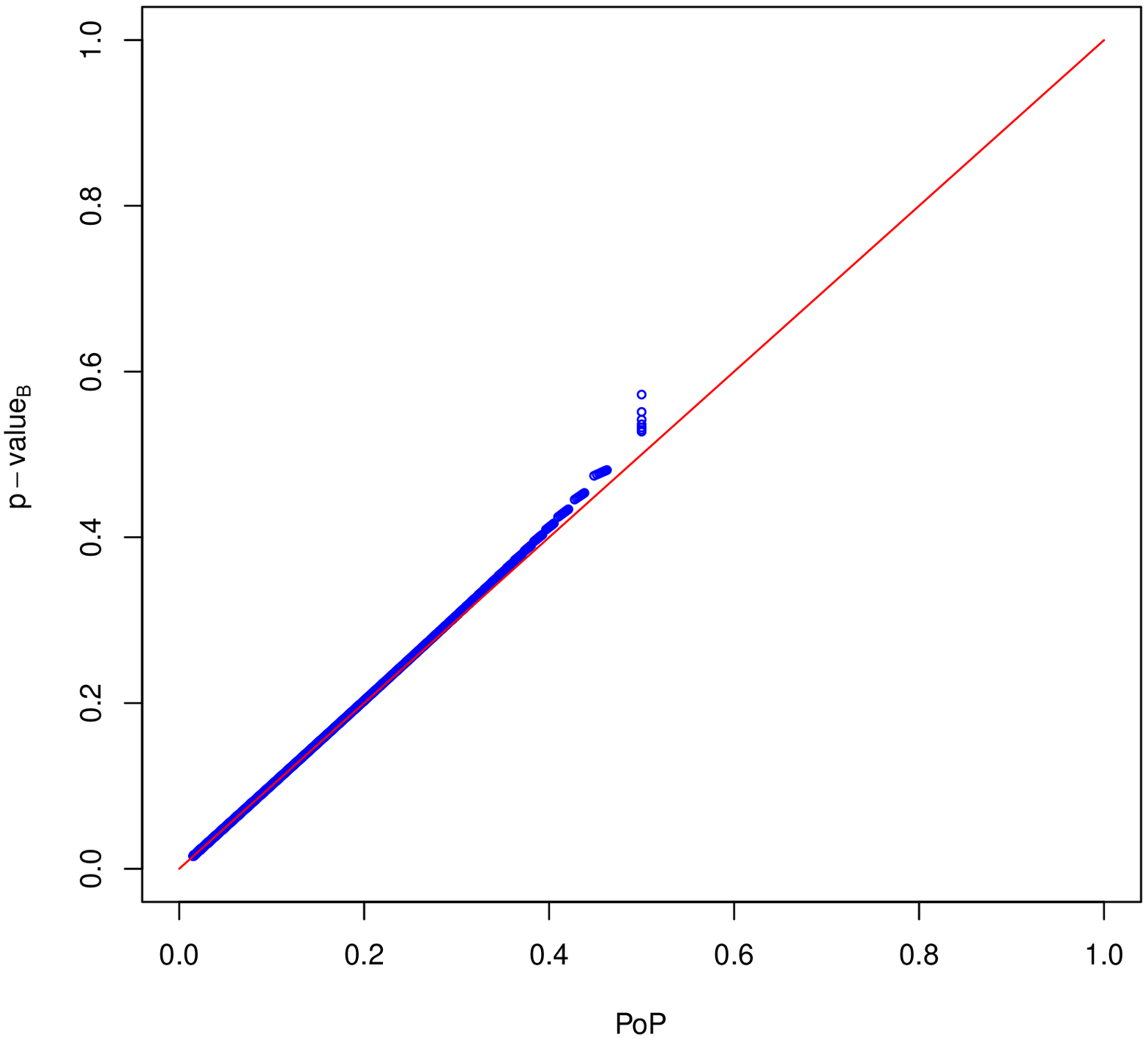}
\includegraphics[angle=0,width=2.7 in, totalheight=2.7 in]{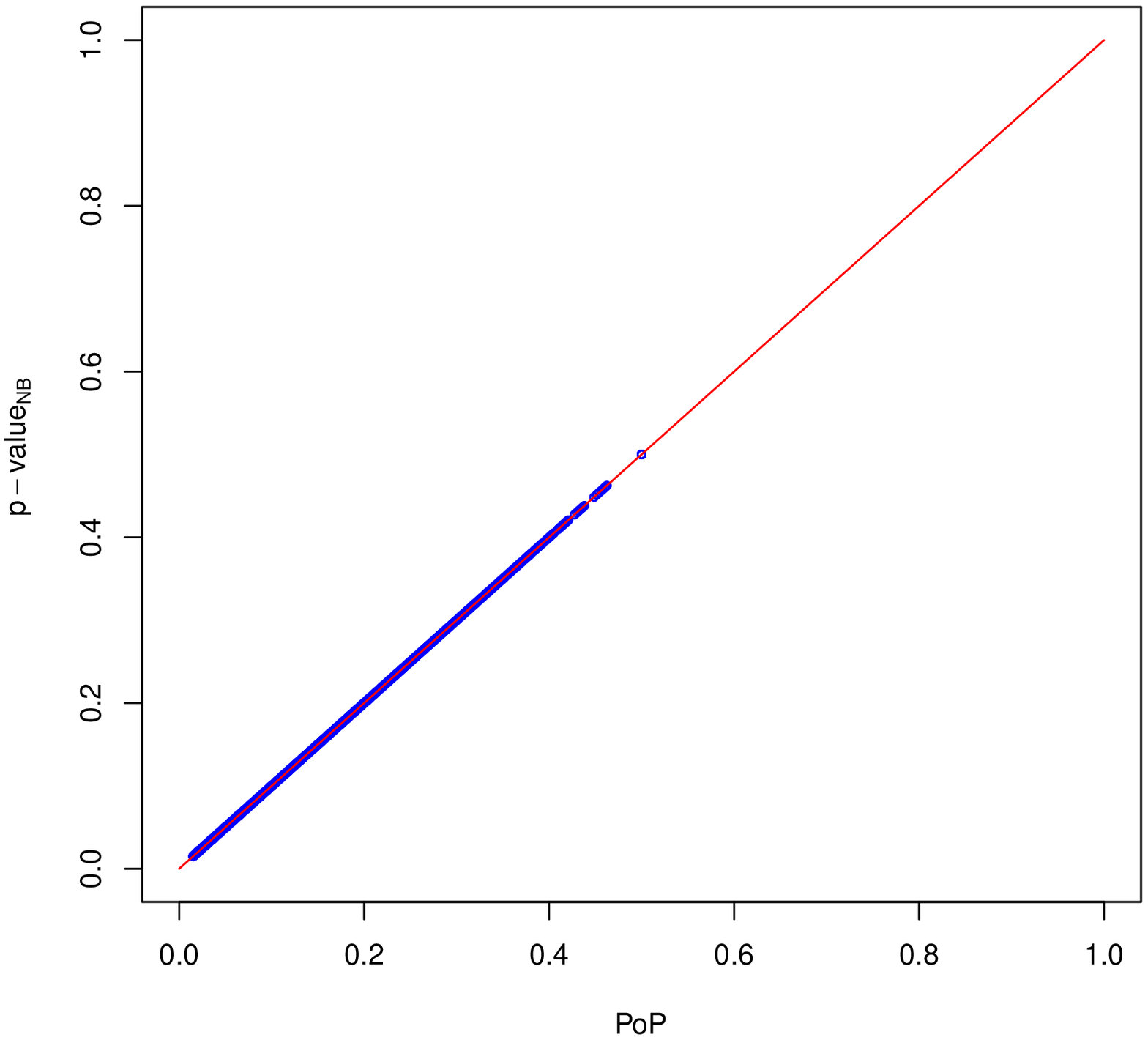}
\end{center}
\caption{The ratio between $p$-values ($p$-value$_{\rm B}$
is based on the binomial distribution, and $p$-value$_{\rm NB}$
is based on the negative binomial distribution,) and the posterior probability (PoP)
of the null hypothesis, as sample size increases while fixing $y/n=0.5044297$.
\label{coinexmple}}
\end{figure}

We further conduct numerical studies to explore the
relationship between the posterior probability of the null hypothesis
and $p$-value. By mimicking the newborn male proportion example,
in the first numerical experiment
we set $y = 0.5044297 \times n$ while increasing $n$ gradually.
In other words, the ratio between $y$ and $n$ is fixed at
the observed value 0.5044297, while
both the values of $y$ and $n$ are increased
to enlarge the sample size. As shown in Figure \ref{coinexmple},
the range of sample size is chosen
such that $p$-values can cover from 0 up to around 0.5. As the
sample size increases, the $p$-value decreases.
It further confirms that the $p$-values under
the negative binomial distribution
match well with the posterior probabilities of $H_0$, while
those under the binomial distribution deviate substantially for all
sample sizes considered.

\begin{figure}[htbp]
\setlength{\abovecaptionskip}{-5pt}
\setlength{\belowcaptionskip}{5pt}
\begin{center}
\includegraphics[angle=0,width=3. in, totalheight=3. in]{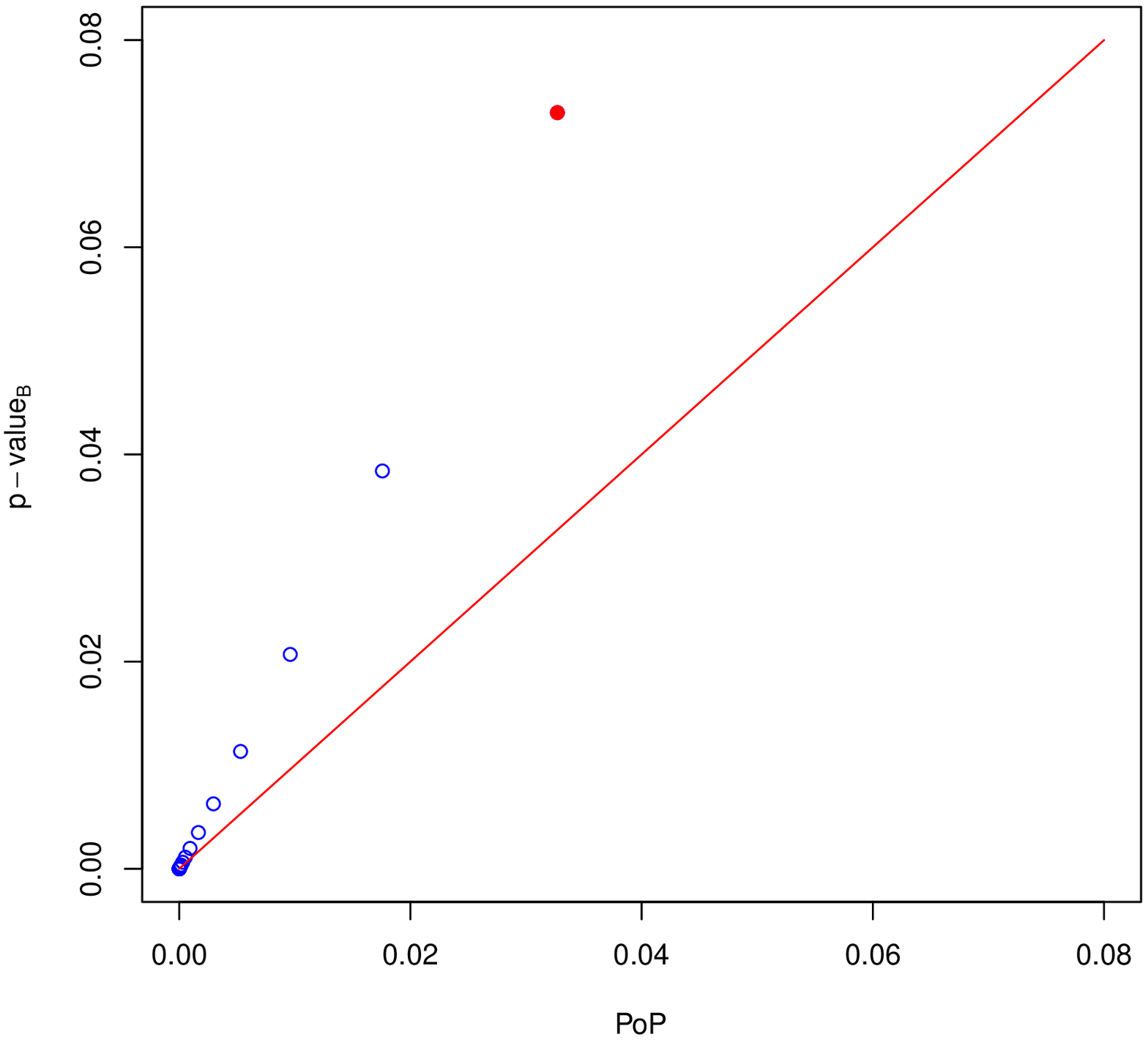}
\includegraphics[angle=0,width=3. in, totalheight=3. in]{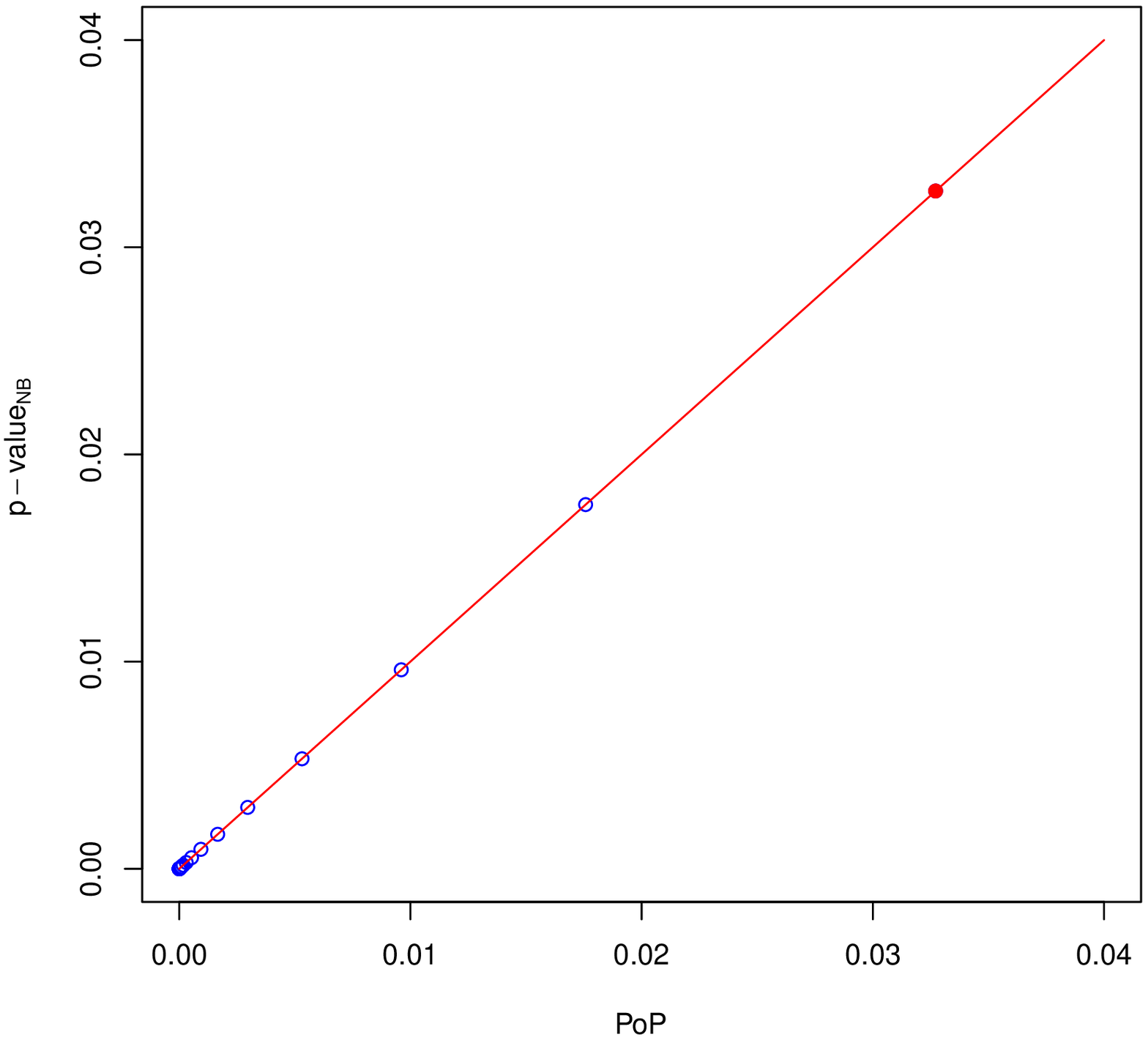}\\
\includegraphics[angle=0,width=3. in, totalheight=3. in]{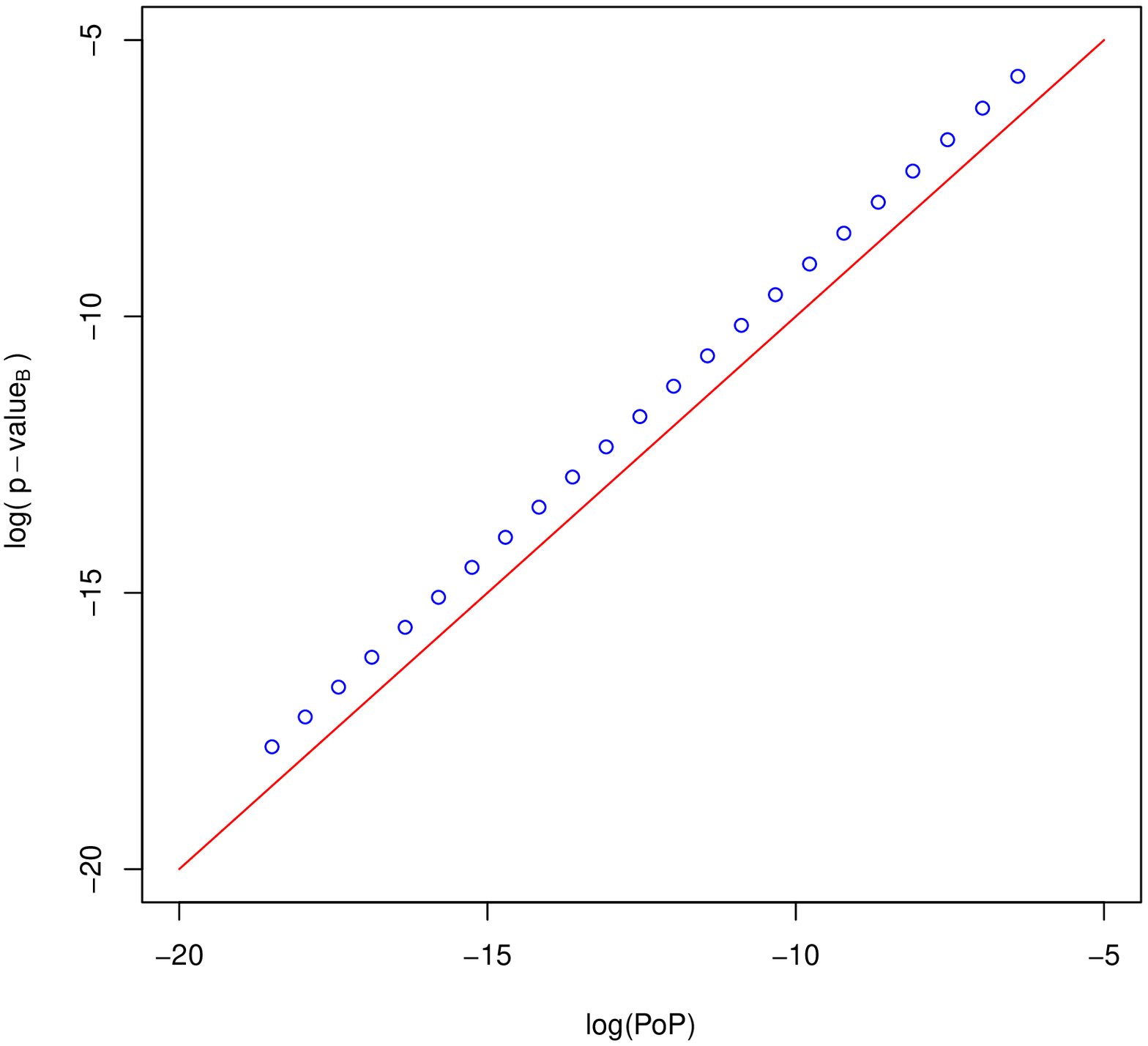}
\includegraphics[angle=0,width=3. in, totalheight=3. in]{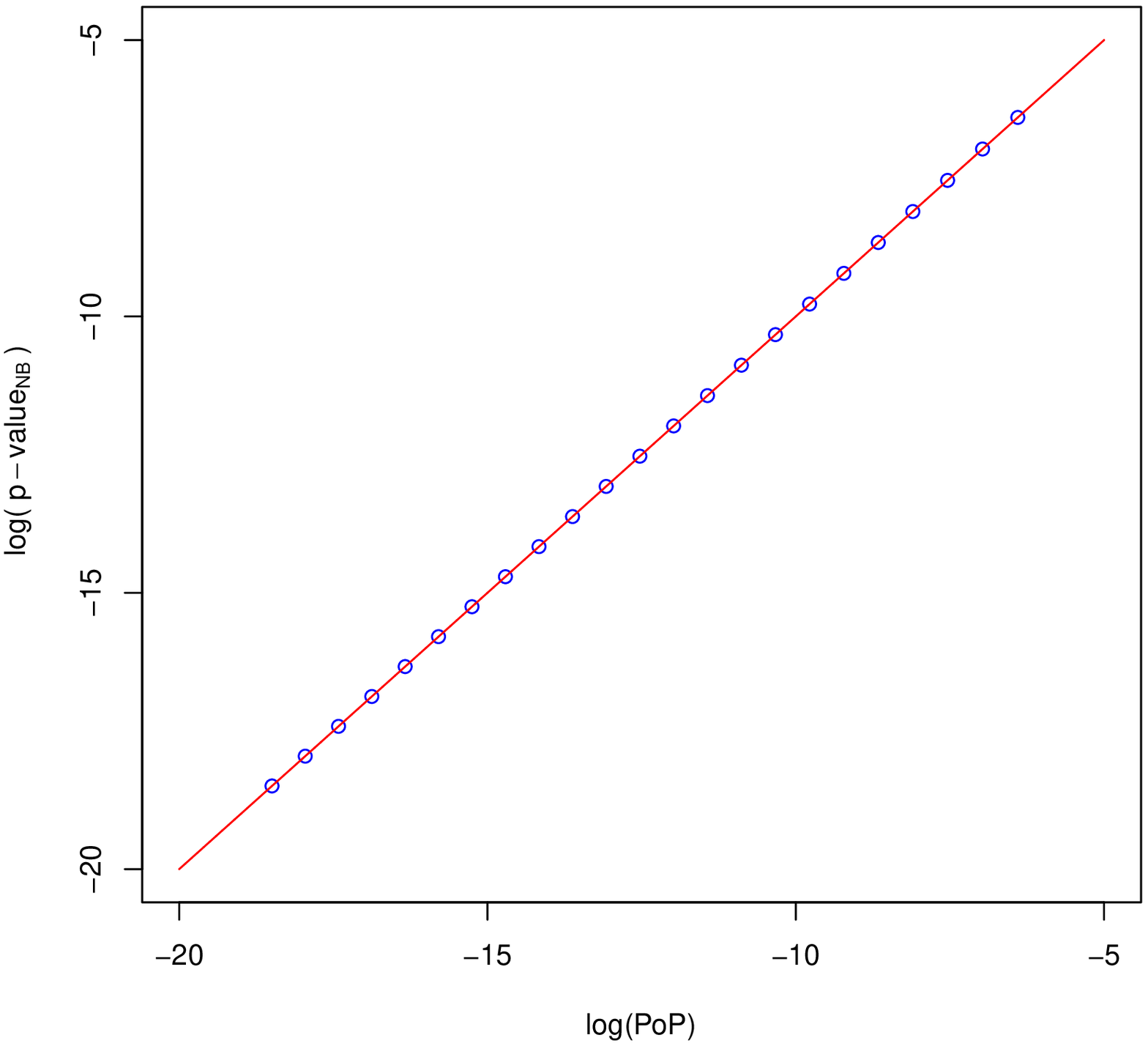}
\end{center}
\caption{The relationship between $p$-values
($p$-value$_{\rm B}$
is based on the binomial distribution, and $p$-value$_{\rm NB}$
is based on the negative binomial distribution,) and the posterior probability (PoP)
of the null
when $y/n$ is fixed at 0.75. The red solid point in the first row corresponds
to the original experiment with $n=12$ and $y=9$.
The second row presents the zoom-in plot at the corner $(0,0)$ of the first row,
i.e., the log-ratio between $p$-values and PoP
for $p$-values smaller than 0.002.
\label{noconv}}
\end{figure}

In the second numerical experiment, we
follow the coin-tossing example by fixing $y/n = 9/12$, while gradually
increasing $n$ up to 120. A non-informative beta prior,
Beta$(10^{-6}, 10^{-6})$, is used. Figure \ref{noconv} again shows that the
$p$-values under the negative binomial distribution
match well with the posterior probability of $H_0$, while those
under the binomial distribution do not.

\section{Hypothesis Tests with Normal Data}
\subsection{Improper Flat Prior}
Consider a two-sample test with normal data.
Let $n$ denote the sample size for each group, and let
$D$ denote the observed data. Assume the outcomes in
groups 1 and 2 to be normally distributed, i.e.,
$y_{1i} \sim {\rm N}(\mu_1, \sigma^2)$ and $y_{2i} \sim {\rm N}(\mu_2, \sigma^2)$
with unknown means $\mu_1$ and $\mu_2$ but a known variance $\sigma^2$ for simplicity.
Let $\bar y_1 = \sum_{i=1}^{n} y_{1i}/n$ and $\bar y_2 = \sum_{i=1}^{n} y_{2i}/n$
be the sample means, and $\theta = \mu_1-\mu_2$ and
$\hat \theta = \bar y_1-\bar y_2$.

\subsubsection{One-sided Test}
We are interested in the one-sided hypothesis test,
\begin{equation*}\label{Hypothesis}
H_0\mbox{:} \ \theta \le 0 \quad  {\rm versus} \quad H_1\mbox{:} \ \theta > 0,
\end{equation*}
the frequentist $Z$-test statistic is formulated as
$$ z=\frac{\bar y_1-\bar y_2}{\sqrt{2\sigma^2/n}}=\frac{\hat \theta}{\sqrt{2\sigma^2/n}},$$
which follows the standard normal distribution under the null hypothesis.
The corresponding $p$-value under the one-sided hypothesis test is given by
\begin{equation}\label{p1exact}
  p \mbox{-value}_1 = \Pr(Z\ge  \hat \theta \sqrt{n/(2\sigma^2)}|H_0)
   = 1 - \Phi( \hat \theta \sqrt{n/(2\sigma^2)}),
\end{equation}
where $Z$ denotes the standard normal random variable and $\Phi(\cdot)$
is the cumulative distribution function of $Z$.

In the Bayesian paradigm, if we assume an improper flat prior distribution,
i.e., $p(\theta) \propto 1$, the posterior distribution of $\theta$ is
$$\theta|D \sim {\rm N}(\hat \theta, 2\sigma^2/n).$$
Therefore, the posterior probability of the null hypothesis is
$$
{\rm PoP}_1 =\Pr(H_0|D)=
\Pr(\theta \le 0 | D) = 1- \Phi( \hat \theta  \sqrt{n/(2\sigma^2)}),
$$
which is exactly the same as (\ref{p1exact}).
Under such an improper flat prior distribution of $\theta$,
we can establish an exact equivalence relationship between
$p$-value and $\Pr(H_0 | D)$.

\subsubsection{Two-sided Test}
Under the two-sided hypothesis test,
\begin{equation*}\label{Hypothesis}
H_0\mbox{:} \ \theta = 0 \quad  {\rm versus} \quad H_1\mbox{:} \ \theta \neq 0,
\end{equation*}
the $p$-value is given by
\begin{eqnarray}\label{p2exact}
 p \mbox{-value}_2 &=& 2[1- {\rm max}\{\Pr(Z\ge z|H_0),\Pr(Z \le z|H_0)\} ]\nonumber\\
   &=& 2 - 2 {\rm max} \{\Phi( \hat \theta \sqrt{n/(2\sigma^2)}),\Phi(- \hat \theta \sqrt{n/(2\sigma^2)})\}.
\end{eqnarray}
The two-sided test can be viewed as a combination of two one-sided
tests (along the opposite directions), and thus the prior distribution
can be easily specified as that in the one-sided test. Otherwise, the
point mass under the null hypothesis poses great challenges for Bayesian
prior specifications.
As a result, the two-sided posterior probability is defined as
\begin{eqnarray*}
{\rm PoP}_2 =\Pr(H_0|D)
&=& 2[1 - {\rm max}\{\Pr(\theta \le 0 | D), \Pr(\theta \ge 0 | D)\}]\\
 &=& 2 - 2 {\rm max} \{\Phi( \hat \theta \sqrt{n/(2\sigma^2)}),\Phi(- \hat \theta \sqrt{n/(2\sigma^2)})\},
\end{eqnarray*}
which is exactly the same as the (two-sided) $p$-value in (\ref{p2exact}).

\subsection{Normal Prior}
\subsubsection{One-sided Test}
If we assume a normal prior distribution for $\theta$, i.e.,
$\theta \sim {\rm N}(\mu_0,\sigma_0^2)$,
the posterior distribution of $\theta$ is still normal,
$\theta|D \sim {\rm N}(\tilde \mu,\tilde \sigma^2)$, where the posterior mean
and the posterior variance are respectively given by
\begin{equation*}
  \tilde \mu = \frac{{\hat \theta} \sigma_0^2+\mu_0(2\sigma^2/n)}{\sigma_0^2  + 2\sigma^2/n},\quad
  \tilde \sigma^2 = \frac{\sigma_0^2(2\sigma^2/n)}{\sigma_0^2  + 2\sigma^2/n}.
\end{equation*}
Under a one-sided test, the posterior probability of $H_0$ is
\begin{eqnarray*}
{\rm PoP}_1=\Pr(H_0 | D) &=& \Pr(\theta \le 0 | D) \\
&=& 1- \Phi({\tilde \mu}/{\tilde \sigma})\\
 &=& 1-\Phi\bigg(\frac{{\hat \theta} \sigma_0^2+\mu_0(2\sigma^2/n)}
 {\sqrt{\sigma_0^2+2\sigma^2/n}}\cdot \frac{1}{\sigma_0\sqrt{2\sigma^2/n}} \bigg)\\
 &=& 1-\Phi\bigg(\frac{{\hat \theta} +\mu_0(2\sigma^2/n)/\sigma_0^2}
 {\sqrt{1+(2\sigma^2/n)/\sigma_0^2}}\cdot \frac{1}{\sqrt{2\sigma^2/n}} \bigg).
\end{eqnarray*}
Therefore, it is evident that as $\sigma_0 \rightarrow \infty$ (i.e., under
non-informative priors),
the posterior probability of the null converges to
\begin{equation*}\label{1side}
{\rm PoP}_1=\Pr(H_0 | D) \approx 1-\Phi( \hat \theta  \sqrt{n/(2\sigma^2)}),
\end{equation*}
which equals the $p$-value under a one-sided hypothesis test. That is,
$
p \mbox{-value}_1  = \lim_{\sigma_0^2 \rightarrow \infty} \Pr(H_0 | D).
$

\subsubsection{Two-sided Test}
For a two-sided hypothesis test, we can also assume a normal prior distribution for $\theta$, i.e.,
$\theta \sim {\rm N}(\mu_0,\sigma_0^2)$, and
the asymptotic equivalence between $p$-value and the posterior probability
of the null can be derived along similar lines.
In particular, we view the two-sided hypothesis test as the combination
of two one-sided tests and $\Pr(\theta \le 0 | D)$ is the same as
(\ref{1side}). For the other one-sided test, as $\sigma_0\to\infty$,
\begin{eqnarray*}
\Pr(\theta \ge 0 | D)
&=& 1- \Phi(-{\tilde \mu}/{\tilde \sigma})\\
 &=& 1-\Phi\bigg(-\frac{{\hat \theta} +\mu_0(2\sigma^2/n)/\sigma_0^2}
 {\sqrt{1+(2\sigma^2/n)/\sigma_0^2}}\cdot \frac{1}{\sqrt{2\sigma^2/n}} \bigg)\\
 &\approx& 1-\Phi( -\hat \theta  \sqrt{n/(2\sigma^2)}).
\end{eqnarray*}
By combining the two one-sided tests,
the two-sided posterior probability is given by
\begin{eqnarray*}
{\rm PoP}_2 =\Pr(H_0|D)
&=& 2[1 - {\rm max}\{\Pr(\theta \le 0 | D), \Pr(\theta \ge 0 | D)\}]\\
 &=& 2 - 2 {\rm max} \{\Phi( \hat \theta \sqrt{n/(2\sigma^2)}),\Phi(- \hat \theta \sqrt{n/(2\sigma^2)})\},
\end{eqnarray*}
which is the same as the (two-sided) $p$-value in (\ref{p2exact}).

\section{Hypothesis Test for Multivariate Normal Data}

In hypothesis testing on the mean vector of a multivariate normal random variable,
we consider $\bX \sim {\rm N}_p(\bmu,\bSigma)$, where $p$ is the dimension of the multivariate normal distribution.
For ease of exposition, the covariance matrix $\bSigma$ is assumed to be known.
Let $D = \{\bX_1,\ldots,\bX_n\}$ denote the observed multivariate vectors, let $\bar \bX = \sum_{i=1}^{n}\bX_i/n$ denote the sample mean vector, and thus $\bar \bX \sim {\rm N}_p(\bmu,\bSigma/n)$.

Consider the one-sided hypothesis test,
\begin{equation*}\label{Hypothesis}
H_0\mbox{:} \ \bc_k^{\top}\bmu \le 0 \ {\rm for} \ {\rm some} \ k = 1,\ldots,K \quad  {\rm versus} \quad H_1\mbox{:} \ \bc_k^{\top}\bmu > 0 \ {\rm for} \ {\rm all} \ k = 1,\ldots,K,
\end{equation*}
where $\bc_1,\ldots,\bc_K$ are $K$ prespecified $p$-dimensional vectors.
The likelihood ratio test statistics (Sasabuchi, 1980) are given by
\begin{equation}\label{sasa}
Z_k =  \frac{\bc_k^{\top} {\bar {\bX}}}{\sqrt{\bc_k^{\top} { \bSigma}\bc_k/n}}, \ \  k = 1,\ldots,K,
\end{equation}
and the corresponding $p$-values are
$$
p \mbox{-value}(k)_1  = 1 - \Phi(Z_k).
$$
The null hypothesis is rejected if all of the $K$ $p$-values are smaller than $\alpha$.

In the Bayesian paradigm, we assume a conjugate multivariate normal prior distribution for $\bmu$,
i.e., $\bmu \sim {\rm N}_p(\bmu_0,\bSigma_0)$.
The corresponding posterior distribution is
$\bmu|D \sim {\rm N}_p(\bmu_n,\bSigma_n)$,
where
\begin{align*}
  \bmu_n & = \bSigma_0\left(\bSigma_0+ \frac{\bSigma}{n}\right)^{-1}{\bar \bX} + \frac{1}{n} \bSigma \left(\bSigma_0+ \frac{\bSigma}{n}\right)^{-1}\bmu_0, \\
  \bSigma_n & = \frac{1}{n}\bSigma \left(\bSigma_0+ \frac{\bSigma}{n}\right)^{-1}\bSigma.
\end{align*}
The one-sided posterior probability corresponding to $\bc_k$ is
$$
{\rm PoP}(k)_1 = \Pr(\bc_k^{\top}\bmu \le 0 | D).
$$

For two-sided hypothesis testing (Liu and Berger, 1995), we are interested in
\begin{eqnarray*}\label{Hypothesis}
&&H_0\mbox{:} \ \bc_k^{\top}\bmu \le 0 \ {\rm for} \ {\rm some} \ k = 1,\ldots,K, {\rm and} \\
&& \ \ \ \ \ \  \bc_k^{\top}\bmu \ge 0 \ {\rm for} \ {\rm some} \ k = 1,\ldots,K \\
&&{\rm versus}  \\
&&H_1\mbox{:} \ \bc_k^{\top}\bmu > 0 \ {\rm for} \ {\rm all} \ k = 1,\ldots,K, \ {\rm or} \\
&& \ \ \ \ \ \ \bc_k^{\top}\bmu < 0 \ {\rm for} \ {\rm all} \ k = 1,\ldots,K.
\end{eqnarray*}
Based on (\ref{sasa}), the $p$-values are given by
$$
p \mbox{-value}(k)_2  = 2 - 2\Phi(|Z_k|) = 2[1-{\rm max}\{ \Phi(Z_k),\Phi(-Z_k) \}].
$$
The null hypothesis is rejected if all of the $K$ $p$-values are smaller than $\alpha$.
Similar to the univariate case, we define the two-sided posterior probability,
$${\rm PoP}(k)_2 = 2[1-{\rm max}\{\Pr(\bc_k^{\top}\bmu > 0 | D), \Pr(\bc_k^{\top}\bmu < 0 | D)\}].$$

\begin{figure}[htbp]
\setlength{\abovecaptionskip}{-5pt}
\setlength{\belowcaptionskip}{5pt}
\begin{center}
\includegraphics[angle=0,width=3. in, totalheight=3. in]{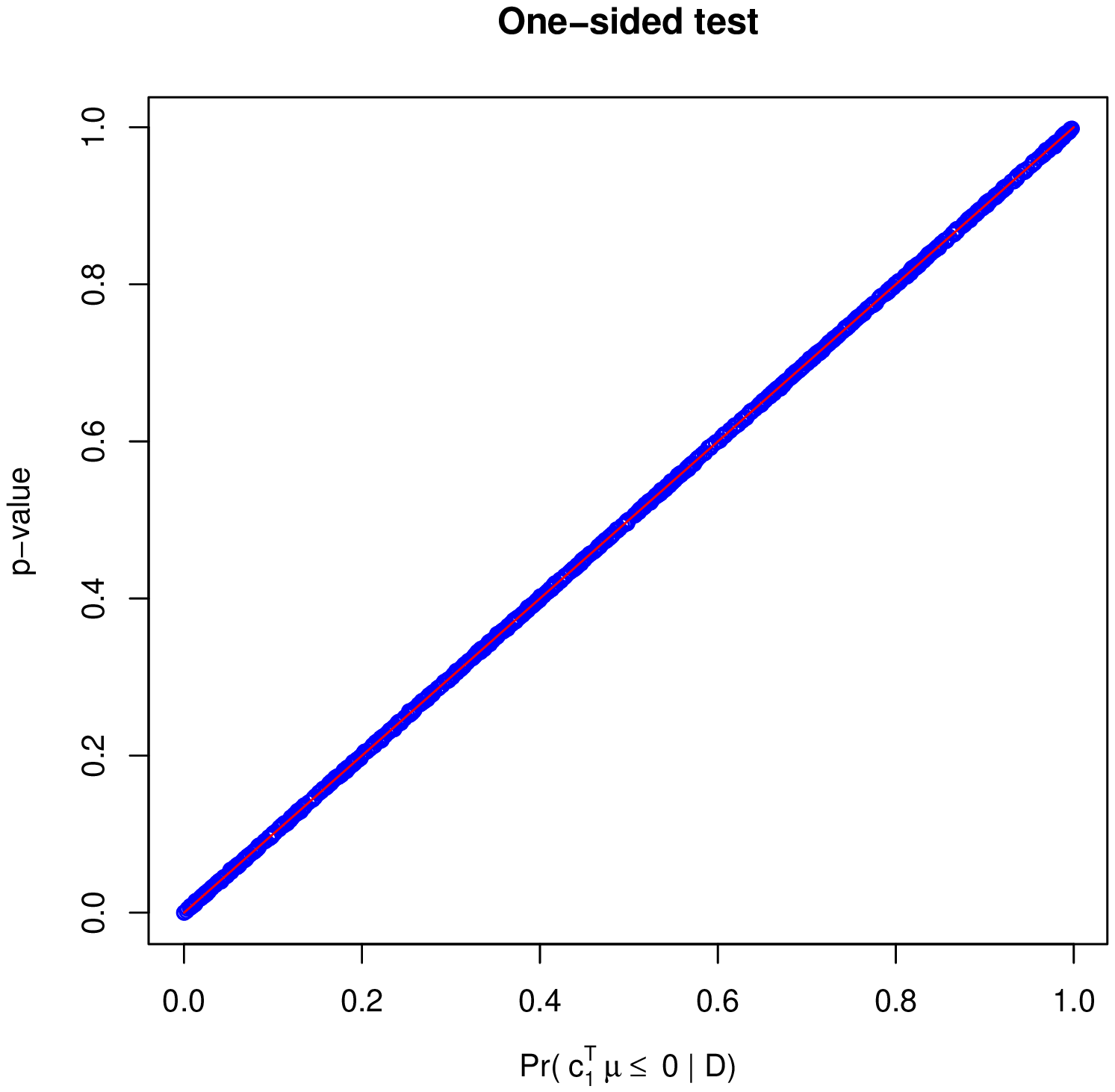}
\includegraphics[angle=0,width=3. in, totalheight=3. in]{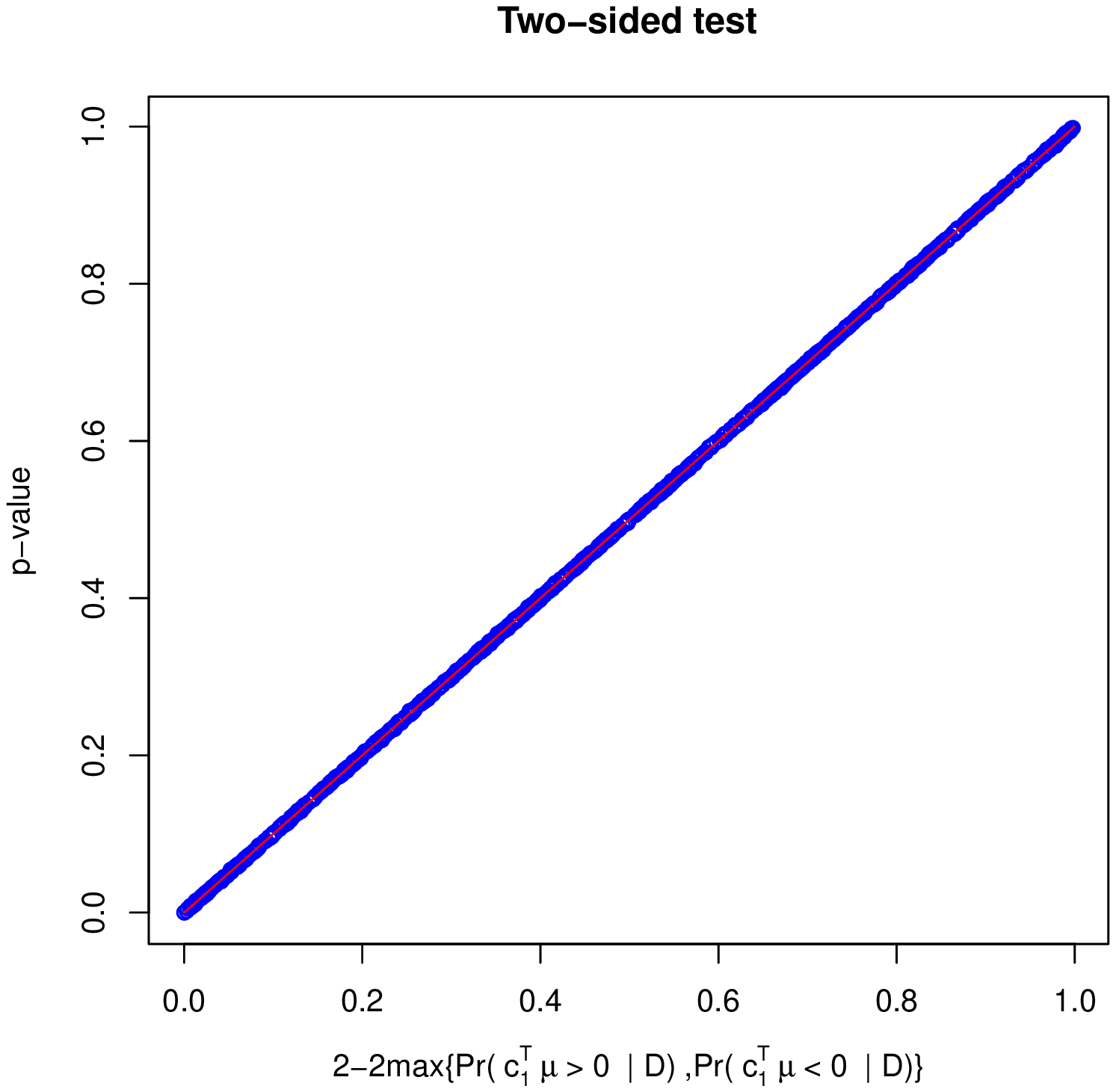}\\
\includegraphics[angle=0,width=3. in, totalheight=3. in]{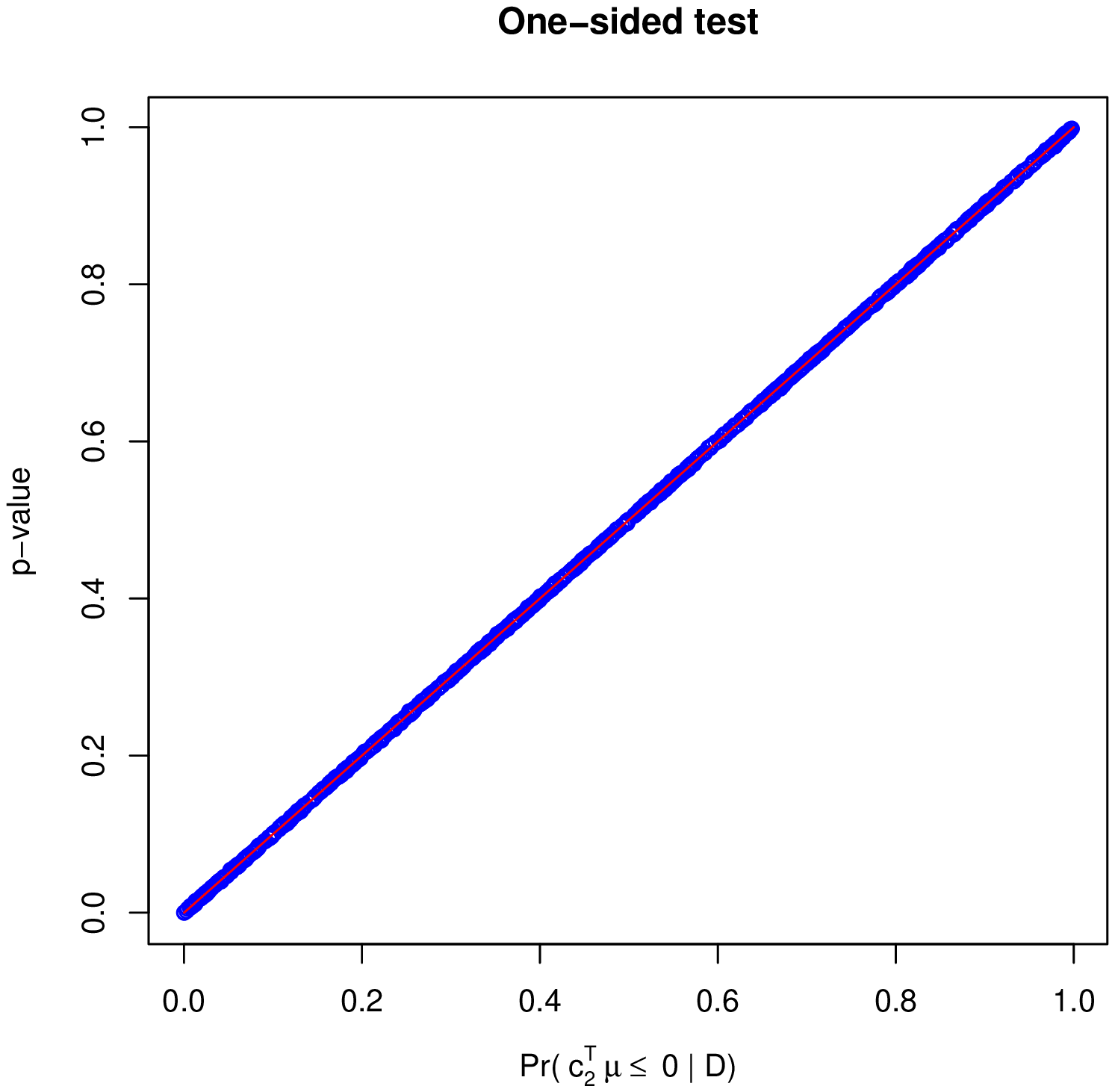}
\includegraphics[angle=0,width=3. in, totalheight=3. in]{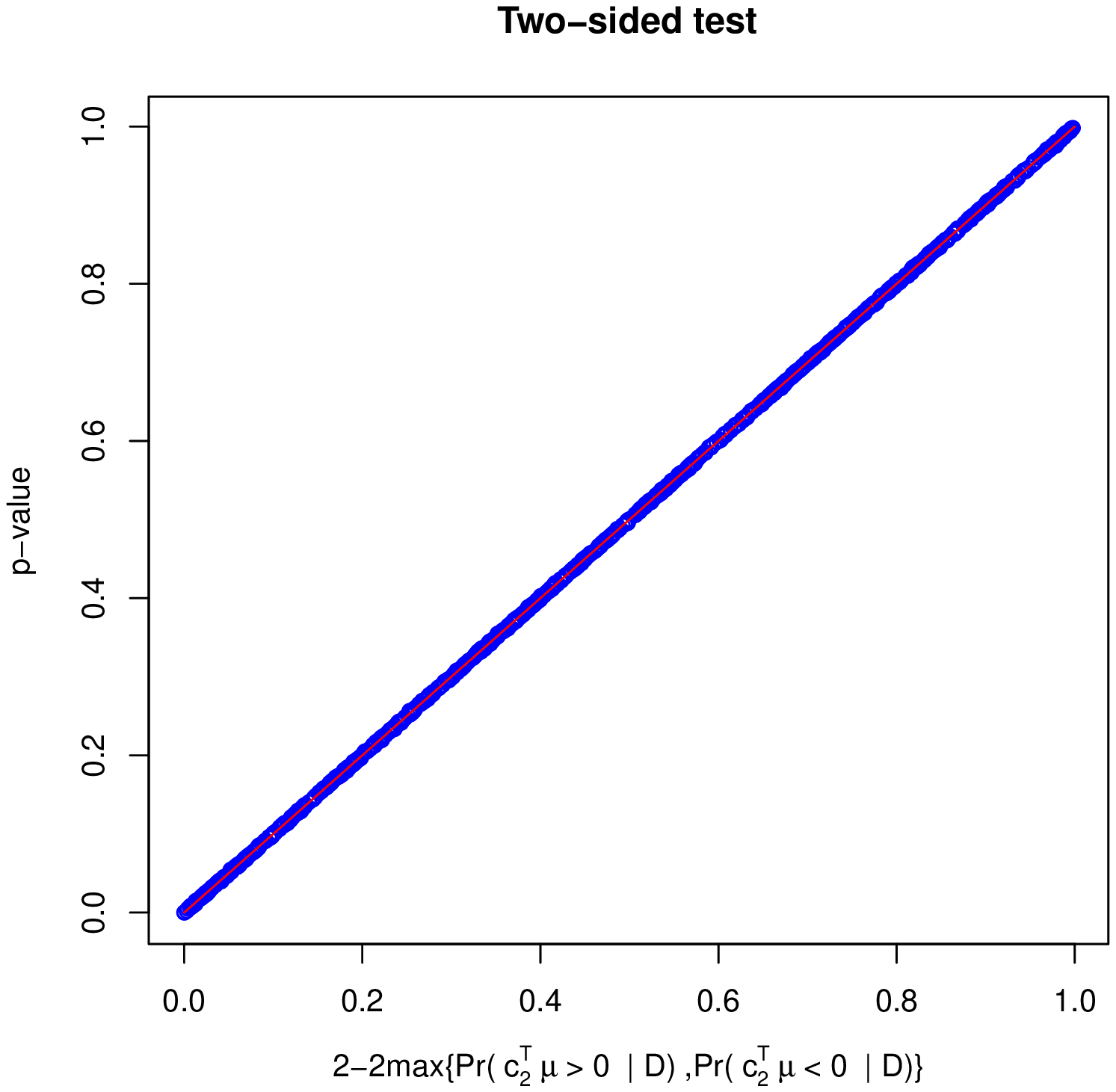}
\end{center}
\caption{The relationship between $p$-value and the posterior probability over 1000 replications under one-sided and two-sided hypothesis tests with multivariate normal outcomes under sample size of 100.
\label{multi}}
\end{figure}

For illustration, we conduct a numerical study to compute the
posterior probabilities of $\bc_k^{\top}\bmu \le 0$ for $k = 1,\ldots,K$, and compare them with the corresponding $p$-values. We take $K=2$ and $\bc_k$ to be a unit vector with 1 on the $k$th element and 0 otherwise, and assume a vague normal prior distribution for $\bmu$, i.e., $\bmu_0 = {\bf 0}$ and $\bSigma_0 = 1000 \bI_p$, where $\bI_p$ is a $p$-dimensional identity matrix.
The relationship between the posterior probabilities of the null
and $p$-values is shown in Figure \ref{multi},
which again demonstrates their equivalence for both one-sided and two-sided tests.

\section{Random Effects Models}
We further consider a random effects model and conduct
hypothesis testing for both regression coefficients and
variance components. The data are
generated from a linear random effects model as follows,
\begin{align*}
y_{ij}=\beta_0+\beta_1x_{ij}+b_{i}+\varepsilon_{ij},
\end{align*}
where $y_{ij}$ is the outcome of observation $j$ in cluster $i$, $i=1,\dots,n$;
$j=1,\dots,J$ and covariates $x_{ij}$'s are generated from ${\rm Unif}(-1,1)$.
We assume $b_i \sim {\rm N}(0,\tau^2)$ and $\varepsilon_{ij} \sim {\rm N}(0,\sigma^2)$.
The sample size is $n=100, 500$, and the cluster size is $J=2, 5$,
and we set the true parameter values to be
$\beta_0=0.2$, $\beta_1=1$ and $\tau=\sigma=0.5$ in our numerical study.

We consider a one-sided test for $\beta_1$,
\begin{align*}
{\rm Test} \ 1: \ \ H_0: \beta_1\leq\delta \quad ~ {\rm versus}\quad ~ H_1:  \beta_1>\delta,
\end{align*}
and another one-sided test for $\tau$,
\begin{align*}
{\rm Test} \ 2: \ \ H_0: \tau^2\leq\xi \quad ~ {\rm versus}\quad ~ H_1: \tau^2>\xi.
\end{align*}
We vary the values of $\delta$ and $\xi$, and for each configuration
we use the Wald test to obtain the $p$-value and compare it with
the posterior probability (PoP) of the null hypothesis.
For the frequentist test on $\tau^2$, we use the asymptotic distribution
based on the Fisher information,
$$\sqrt{N}(\hat{\tau}^2 - \tau^2) \xrightarrow{\cal D}
{\rm N}\left(0,\frac{2\sigma^4}{J(J-1)}+\frac{2(J\tau^2+\sigma^2)^2}{J}\right),$$
and by the delta method, we take the log transformation,
$$\sqrt{N}(\log(\hat{\tau}^2) - \log(\tau^2)) \xrightarrow{\cal D}
{\rm N}
\left(0,\left(\frac{2\sigma^4}{J(J-1)}
+\frac{2(J\tau^2+\sigma^2)^2}{J}\right)\frac{1}{\tau^4}\right).$$

\begin{figure}[htbp]
\begin{center}
\includegraphics[width=\columnwidth]{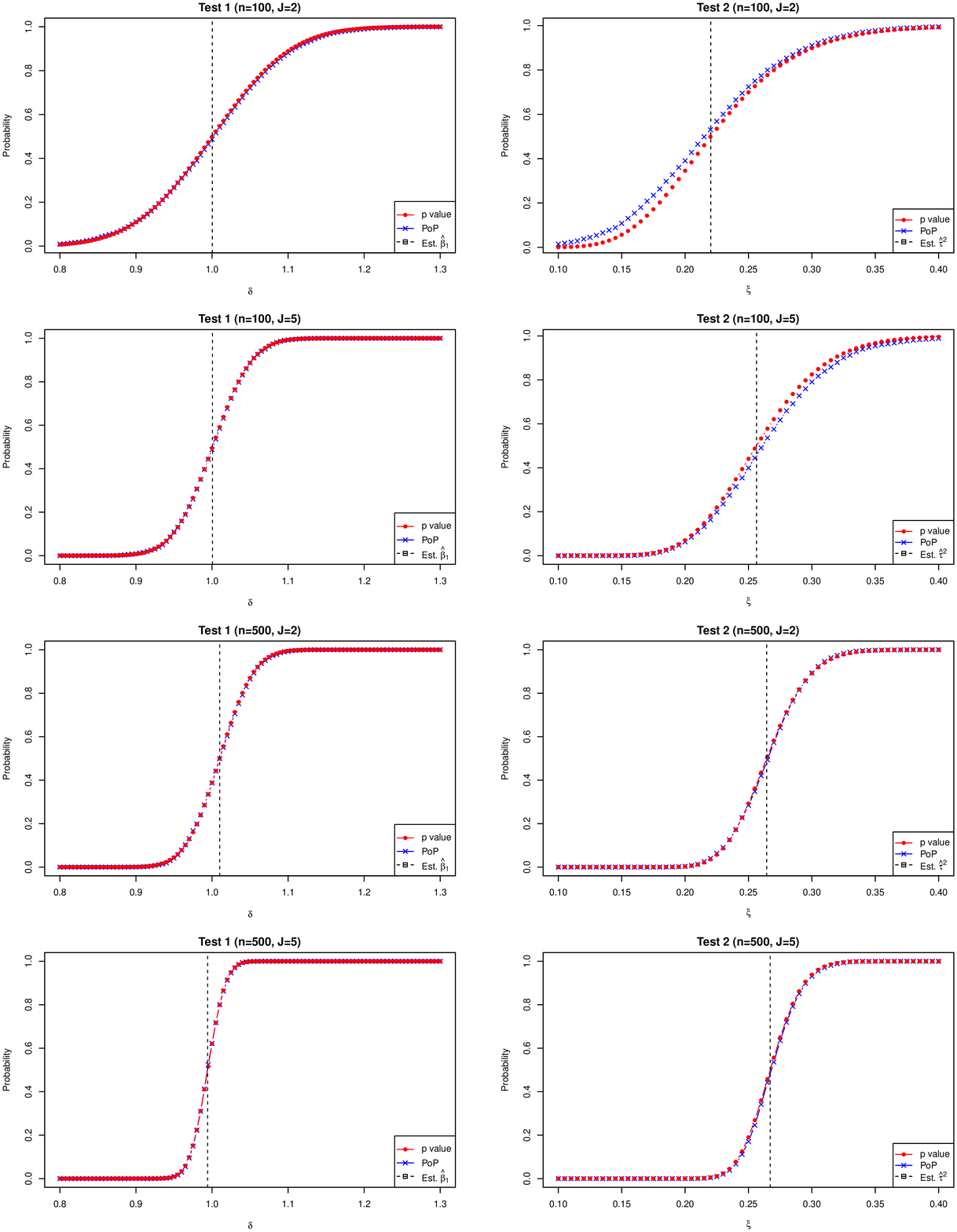}
\caption{P-values and posterior probabilities of the null under
 the tests 1 and 2 with $n = (100,500),~J=(2,5)$.}
\label{figRanEff}
\end{center}
\end{figure}

Figure \ref{figRanEff} shows that for different values of $\delta$
and $\xi$ the $p$-values and the posterior probabilities of the null hypothesis
are very close, especially under the settings of $n=500$. The match
between the two quantities appear to be better for the tests of
the regression coefficient than those of the variance component.

\section{Discussion}
The $p$-value is the most commonly used summary measure for evidence-based studies, and it
has been the center of controversies and debates for decades.
Recently reignited discussion over $p$-values has been more centered 
around the proposals to adjust, abandon or provide alternatives to $p$-values.
By definition, $p$-value is not the probability
that the null hypothesis is true given the observed data.
Contrary to the conventional notion,
it does have a close correspondence to the Bayesian posterior 
probability of the null hypothesis being true
for both one-sided and two-sided hypothesis tests.
Certainly, such equivalence relationship would not
hold when informative priors are used, because $p$-values are computed
without any prior information involved.
Lindley's paradox mainly
arises when a point mass is put on the parameter of interest under
the null hypothesis. We circumvent the controversy by recasting a two-sided
hypothesis into two one-sided hypotheses, and then the paradox can be explained:
the $p$-value and
the Bayesian posterior probability of the null hypothesis coincide.

\section*{References}
\begin{description}

\item
Bayarri, M. J. and Berger, J. O. (2004). The interplay of Bayesian and frequentist analysis. {\em Statistical Science} {\bf 19}, 58--80.

\item
Benjamin, D. J. and Berger J. O. (2019). Three recommendations for improving the use of p-values. {\em The American Statistician} {\bf 73}, 186--191.

\item
Berger, J. O. (2003). Could Fisher, Jeffreys and Neyman have agreed on testing? (with discussion) {\em Statistical Science
}  {\bf 18}, 1--32.

\item
Berger, J. O. and Delampady M. (1987). Testing precise hypotheses. {\em Statistical Science} {\bf 2}, 317--335.

\item
Berger, J. O. and Sellke, T. (1987). Testing a point null hypothesis: the irreconcilability of P values and evidence. {\em Journal of the  American Statistical Association} {\bf 82}, 112--122.


\item
Betensky, R. A. (2019). The p-value requires context, not a threshold. {\em The American Statistician} {\bf 73}, 115--117.

\item
Billheimer, D. (2019). Predictive inference and scientific reproducibility. {\em The American Statistician} {\bf 73}, 291--295.

\item
Casella, G. and Berger, R. L. (1987). Reconciling Bayesian and frequentist evidence in the one-sided testing problem. (with discussion) {\em Journal of the  American Statistical Association} {\bf 82}, 106--111.

\item
Colquhoun, D. (2019). The false positive risk: a proposal concerning what to do about p-values. {\em The American Statistician} {\bf 73}, 192--201.




\item
Gannon, M. A., Pereira, C. A. B., Polpo, A. (2019). Blending Bayesian and classical tools to define optimal sample-size-dependent significance levels. {\em The American Statistician} {\bf 73}, 213--222.

\item
Goodman, S. N. (1999). Toward evidence-based medical statistics. 1: the p value fallacy. {\em Annals of Internal Medicine} {\bf 130}, 995--1004.

\item
Hubbard, R. and Lindsay, R. M. (2008). Why P values are not a useful measure of evidence in statistical significance testing. {\em Theory $\&$ Psychology} {\bf 18}: 69--88.



\item
Ioannidis, J. P. (2019). What have we (not) learnt from millions of scientific papers with p values? {\em The American Statistician} {\bf 73},  20--25.





\item
Johnson, V. E. and Rossell, D. (2010). On the use of non-local prior densities in Bayesian hypothesis tests.
\emph{Journal of the Royal Statistical Society: Series B (Statistical Methodology)} {\bf 72}, 143--170.

\item
Johnson, V. E. (2019). Evidence from marginally significant t statistics. {\em The American Statistician} {\bf 73}, 129--134.

\item
Kennedy-Shaffer, L. (2019). Before p $< 0.05$ to beyond p $< 0.05$: using history to contextualize p-values and significance testing. {\em The American Statistician} {\bf 73}, 82--90.

\item
Krueger, J. I. and Heck, P. R. (2019). Putting the p-value in its place. {\em The American Statistician} {\bf 73}, 122--128.


\item
Lindley, D. V. (1957). A statistical paradox. {\em Biometrika}  {\bf 44}, 187--192.


\item
Manski, C. F. (2019). Treatment choice with trial data: statistical decision theory should supplant hypothesis testing. {\em The American Statistician} {\bf 73}, 296--304.

\item
Matthews, R. A. J.  (2019). Moving towards the post p $< 0.05$ era via the analysis of credibility. {\em The American Statistician} {\bf 73}, 202--212.

\item
McShane, B. B., Gal, D., Gelman, A., Robert, C., and Tackett, J. L. (2019). Abandon statistical significance. {\em The American Statistician} {\bf 73}, 235--245.



\item
Pratt, J. W. (1965). Bayesian interpretation of standard inference statements
(with discussion). {\em Journal of the Royal Statistical Society, Series
B} {\bf 27}, 169--203.


\item
Rougier, J. (2019). p-values, Bayes factors, and sufficiency. {\em The American Statistician} {\bf 73}, 148--151.


%

\item
Ruberg, S. J., Harrell Jr., F. E., Gamalo-Siebers, M., LaVange, L., Lee, J. J., Price, K., Peck, C. (2019). Inference and decision making for 21st-century drug development and approval. {\em The American Statistician} {\bf 73}, 319--327.




\item
Sellke, T., Bayarri, M. J., and Berger, J. O. (2001). Calibration of p-values for testing precise null hypotheses. {\em The American Statistician} {\bf 55}, 62--71.

\item
Shi, H. and Yin, G. (2020).
Reconnecting p-value and posterior probability under one- and two-sided tests.
{\em The American Statistician} DOI: 10.1080/00031305.2020.1717621

\item
Wagenmakers, E. J. (2007). A practical solution to the pervasive problems of p-values.
{\em Psychonomic Bulletin $\&$ Review} {\bf 14}, 779--804.

\item
 Wasserstein, R. L. and Lazar, N. A. (2016). The ASA's statement on p-values: context, process, and purpose. {\em The American Statistician} {\bf 70}, 129--133.

\item
Wasserstein, R. L., Schirm, A. L., and Lazar, N. A. (2019).
Moving to a world beyond ``p$<0.05$''. {\em The American Statistician} {\bf 73},  1--19.
\end{description}

\newpage
\begin{table}[t]
  \centering
  \caption{Relationship between the posterior probability of the null
  hypothesis, $P(H_0|y)$, and
  the values of the hyperparameters in the Beta$(\alpha,\beta)$ prior distribution with
  $\alpha=\beta$ under the Bayesian exact beta posterior distribution.
  \label{popnull}}
\begin{tabular}{lr}
\toprule
$\alpha=\beta$ & $P(H_0|y)$ \\
\midrule
1 & 0.01793728 \\
0.1   & 0.01793580 \\
0.01  & 0.01793565 \\
0.001 & 0.01793564 \\
0.0001 & 0.01793563 \\
0.00001 & 0.01793563 \\
0.000001 & 0.01793563 \\
\bottomrule
\end{tabular}%
  \label{tbl1}%
\end{table}%

\begin{table}[htbp]
  \centering
  \caption{Relationship between the posterior probability $P(H_0|y)$ and
  the values of the hyperparameters in the Beta prior distribution
  $\alpha=\beta$.
  \label{popnull}}
\begin{tabular}{lr}
\toprule
$\alpha=\beta$ & $P(H_0|y)$ \\
\midrule
2     & 0.059235 \\
1.5   & 0.052752 \\
1     & 0.046143 \\
0.9   & 0.044809 \\
0.8   & 0.043471 \\
0.7   & 0.042131 \\
0.6   & 0.040789 \\
0.5   & 0.039445 \\
0.4   & 0.038099 \\
0.3   & 0.036753 \\
0.2   & 0.035406 \\
0.1   & 0.034060 \\
0.01  & 0.032849 \\
0.001 & 0.032728 \\
0.0001 & 0.032716 \\
0.00001 & 0.032715 \\
0.000001 & 0.032715 \\
\bottomrule
\end{tabular}%
  \label{tab:addlabel}%
\end{table}%

\newpage

\newpage

\newpage

\newpage

\newpage

\end{document}